\documentclass[%
 reprint,
superscriptaddress,
 amsmath,amssymb,
 aps,
]{revtex4-2}

\usepackage{graphicx}% Include figure files
\usepackage{dcolumn}% Align table columns on decimal point
\usepackage{bm}% bold math
\usepackage{qcircuit}
\usepackage{physics}
\usepackage{mathtools}
\usepackage{xcolor}
\usepackage{booktabs}
\usepackage{multirow}

\newcommand{\jl}[1]{{\color{red}{#1}}}

\begin{document}

\preprint{APS/123-QED}

\title{Noise-Assisted Digital Quantum Simulation of Open Systems}% Force line breaks with \\

\author{José D. Guimarães}

\affiliation{%
Institute of Theoretical Physics and IQST, Ulm University, Albert-Einstein-Allee 11 89081, Ulm, Germany.
}%

\affiliation{%
Centro de Física das Universidades do Minho e do Porto, Braga 4710-057, Portugal
}%

\affiliation{%
Intl. Iberian Nanotechnology Laboratory,
Av. Mestre Jos{\'e} Veiga s/n, Braga 4715-330, Portugal.
}%
\author{James Lim} 

\affiliation{%
 Institute of Theoretical Physics and IQST, Ulm University, Albert-Einstein-Allee 11 89081, Ulm, Germany.
}%
\author{Mikhail I. Vasilevskiy}

\affiliation{%
Centro de Física das Universidades do Minho e do Porto, Braga 4710-057, Portugal
}%

\affiliation{%
Intl. Iberian Nanotechnology Laboratory,
Av. Mestre Jos{\'e} Veiga s/n, Braga 4715-330, Portugal.
}%
\author{Susana F. Huelga}

\affiliation{%
Institute of Theoretical Physics and IQST, Ulm University, Albert-Einstein-Allee 11 89081, Ulm, Germany.
}%
\author{Martin B. Plenio}

\affiliation{%
Institute of Theoretical Physics and IQST, Ulm University, Albert-Einstein-Allee 11 89081, Ulm, Germany.
}%

\begin{abstract}
%Quantum systems are inherently open and subject to environmental noise, which can have both detrimental and beneficial effects on their dynamics. In particular, noise has been observed to enable novel functionalities in bio-molecular systems, making the simulation of their dynamics an important target for digital and analog quantum simulation. However, current quantum devices are typically noisy, limiting their computational capabilities.
%
%In this work, we propose a novel approach that leverages the intrinsic noise of a quantum device to reduce the quantum computational resources required for simulating open quantum systems. We achieve this by combining quantum noise characterization methods with quantum error mitigation techniques, which allow us to transform and control the intrinsic noise in a quantum circuit. Specifically, we selectively enhance or reduce decoherence rates in the quantum circuit to achieve the desired simulation of open system dynamics.
%
%We describe our methods in detail and report on the results of noise characterization and quantum error mitigation on real and emulated IBM Quantum computers. We also provide estimates of the experimental resource requirements for our techniques. We believe that this approach can pave the way for new simulation techniques in Noisy Intermediate-Scale Quantum (NISQ) devices, where their intrinsic noise can be harnessed to assist quantum computations.

Quantum systems are inherently open and susceptible to environmental noise, which can have both detrimental and beneficial effects on their dynamics. This phenomenon has been observed in bio-molecular systems, where noise enables novel functionalities, making the simulation of their dynamics a crucial target for digital and analog quantum simulation. Nevertheless, the computational capabilities of current quantum devices are often limited due to their inherent noise.
In this work, we present a novel approach that capitalizes on the intrinsic noise of quantum devices to reduce the computational resources required for simulating open quantum systems. Our approach combines quantum noise characterization methods with quantum error mitigation techniques, enabling us to manipulate and control the intrinsic noise in a quantum circuit. Specifically, we selectively enhance or reduce decoherence rates in the quantum circuit to achieve the desired simulation of open system dynamics.
We provide a detailed description of our methods and report on the results of noise characterization and quantum error mitigation experiments conducted on both real and emulated IBM Quantum computers. Additionally, we estimate the experimental resource requirements for our techniques. Our approach holds the potential to unlock new simulation techniques in Noisy Intermediate-Scale Quantum (NISQ) devices, harnessing their intrinsic noise to enhance quantum computations.
\end{abstract}

\maketitle

\section{Introduction}
The dynamics of a closed quantum system in complete isolation from its surroundings is governed
by the Schr{\"o}dinger equation of the system degrees of freedom only and, due to the properties 
of the Hilbert space description of quantum many particle systems, displays a level of complexity
that renders its simulation on a classical computer inefficient. Thus, the founding fathers of 
the field of quantum computation posited that a programmable but otherwise closed quantum device, 
a quantum computer, should be able to simulate this dynamics by harnessing the increased complexity 
of controlled quantum systems to our advantage~\cite{yuri1980computable,feynman1982simulating}. 
Thanks to the sustained research and development effort over the last 30 years we are now reaching 
a situation where the construction of ever more complex quantum information processors is becoming 
a reality. However, in practice interactions with uncontrolled environmental degrees of freedom, a.k.a. 
noise, are unavoidable and render present day quantum information processors unsuitable in principle 
for the efficient simulation of the pure state dynamics of isolated quantum systems. In fact, they are 
so noisy that even quantum error correction with an arbitrary overhead in physical resources would 
not yet be able to remove the effect of this noise. In fact, attempting to apply quantum error 
correction under these conditions would only worsen the overall performance.

Under the influence of intense noise, classical correlations tend to prevail, allowing for an accurate 
and efficient classical description. But for a moderate level of noise, such as that present in today's 
quantum information processors, the situation is different because the full density matrix formalism 
needs to be adopted to achieve an accurate description of the open quantum system. 
While the effect of some environments may be captured accurately by Markovian master
equations~\cite{breuer2002theory,rivas2012open}, the situation is even more challenging for complex 
environments that display temporal correlations as then environmental degrees of freedom need to be
accounted for in some detail making the full system extremely large and challenging to simulate~\cite{tanimura1989time,prior2010efficient,strathearn2018efficient,tamascelli2019efficient,somoza2019dissipation,somoza2022driving,guimaraes2022efficient}.

This added complexity suggests that under suitable conditions and control, environmental noise on quantum 
systems may confer additional benefits, thus motivating the development of methods to simulate 
the properties of noisy quantum systems. The effect of environmental degrees of freedom on the static and 
dynamical properties of a quantum system is especially important in cases, where the time and energy scales 
involved are likely to make interactions between the system and the surrounding environment a key factor in
their own right in the physics at play. Indeed, it was recognised early on in the development of quantum 
information processing that noise and dissipation may serve as a resource for robust entangled 
state preparation~\cite{plenio1999cavity,plenio2002entangled}. More recently, the recognition that quantum dynamics may play
an important role in certain processes of life~\cite{mohseni2014quantum} led to the discovery that the
interplay of coherent quantum dynamics and environmental noise has the potential to impart fundamental
advantages upon these processes~\cite{plenio2008dephasing,mohseni2008environment,caruso2009highly,huelga2013vibrations}.
The electronic properties of bio-molecular complexes are exceedingly difficult to compute and it is 
well-recognized that such quantum chemistry challenges represent a promising application for 
quantum computers~\cite{mcardle2020quantum}. However, these algorithms typically treat closed static systems 
while capturing the properties of environmentally assisted quantum dynamics requires the extension of these
methods to open systems dynamics on a quantum information processor. This implies considerable, potentially 
crippling, overheads when pursued in the standard approaches using fully error corrected and thus noise-free
devices. 

\begin{figure*}
\centering
\includegraphics[width=0.9\textwidth]{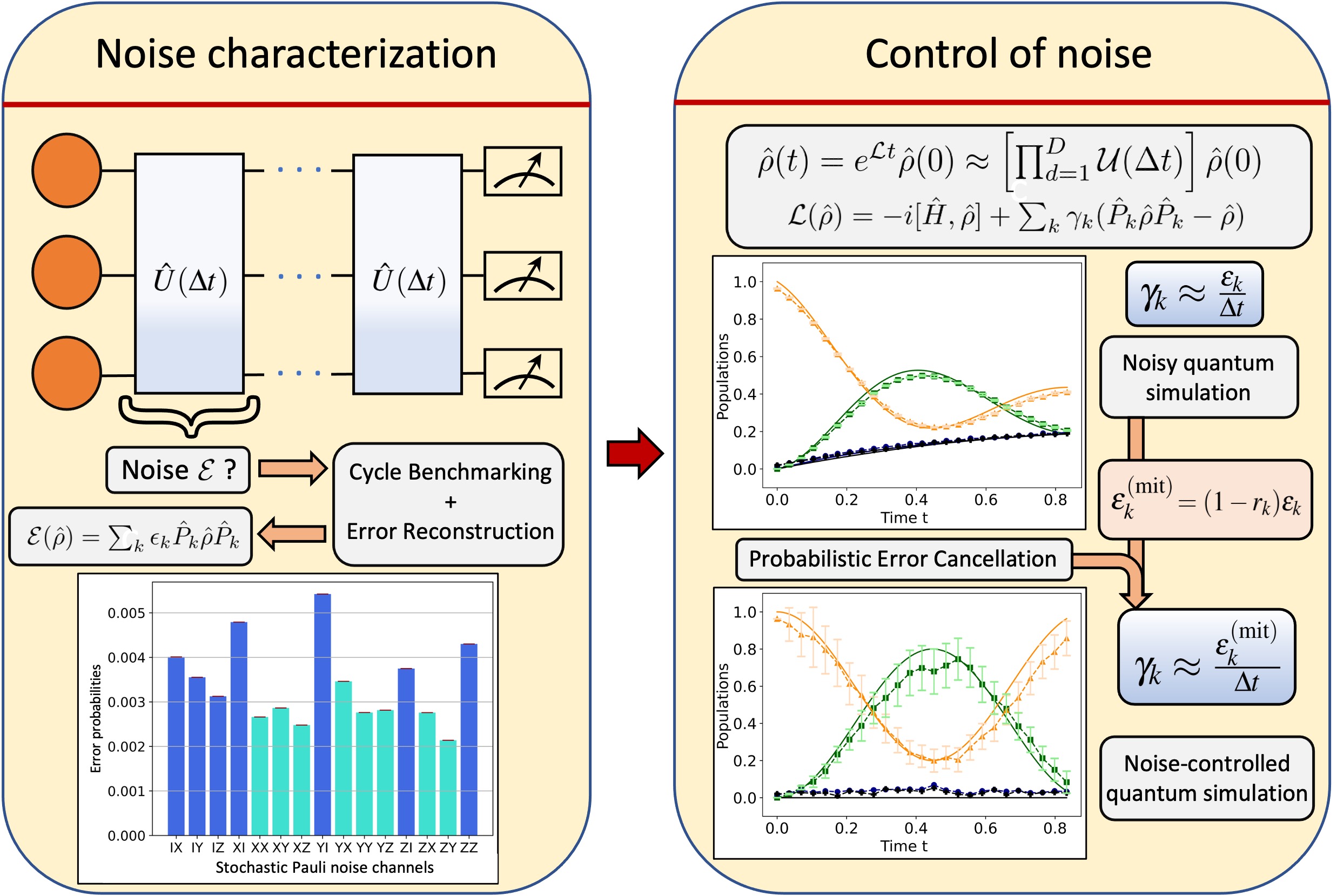}
\caption{Overview of the noise-assisted technique for digital quantum simulation of open system dynamics proposed in this work. The $k$-th order Trotter-Suzuki decomposition of a time-evolution operator ${\rm exp}(-i\hat{H}t)$ is implemented on a quantum circuit for a given open-system Hamiltonian $\hat{H}$, leading to $D$ Trotter layers $\hat{U}(\Delta t)$ with a Trotter time-step $\Delta t=t/D$. The intrinsic noise of quantum devices is transformed to stochastic Pauli noise channels ${\cal E}$ by using Randomized Compiling. In the noise characterization step, the Pauli error probabilities $\epsilon_k$ are estimated once for a noisy Trotter layer ${\cal U}(\Delta t)$ by using Cycle Benchmarking and Error Reconstruction techniques. In the digital quantum simulation of open system dynamics implemented by $D$ Trotter layers, the stochastic Pauli noise induces decoherence effects on the open system dynamics. In the noise control step, the stochastic Pauli noise is partially mitigated in a controlled manner by using a Probabilistic Error Cancellation (PEC) technique, leading to reduced error probabilities $\epsilon_{k}^{\rm (mit)}=\epsilon_k(1-r_k)$ with mitigation factors $r_k\in[0,1]$. With an optimal choice of the Trotter time-step $\Delta t$, this enables one to implement a Lindblad equation with target decoherence rates $\gamma_k = \epsilon_{k}^{\rm (mit)}/\Delta t$ on NISQ devices.}
\label{fig:overview}
\end{figure*}

These considerations raise the natural question whether it might be advantageous to use the presence of
internal noise sources of a quantum information processor in order to increase its efficiency in simulating
a desired open system dynamics. Indeed, very first steps in this direction have been taken in analog quantum
simulators~\cite{lemmer2018trapped} where noise increases the efficiency of the quantum simulation of open 
system dynamics compared to standard approaches~\cite{porras2008mesoscopic}. Nevertheless, until now the 
field of digital quantum computation and simulation has been regarded noise as \textit{detrimental}, to be
fought by quantum error correction, and the question whether it can be leveraged for a particular quantum 
application has not been considered nor have
methods been developed to achieve it. Here we address this challenge by combining the capability of a 
Noisy Intermediate-Scale Quantum (NISQ) computer to execute noisy quantum gate sequences with methods from 
error mitigation and controlled addition of errors to reshape the noise intrinsic to the NISQ device in such a
manner that it models accurately the desired environmental noise of an open quantum system model 
of practical interest (see Fig.~\ref{fig:overview} for an overview of our scheme).

As a major benefit of this approach, the unavoidable decoherence of the quantum bits and quantum gates in a
NISQ device is used to replace some of the quantum computational resources, i.e. qubits and two-qubit quantum 
gates, that would otherwise be necessary to simulate faithfully the effect of noise in the desired open quantum
system model. Thus, the overall resource requirements for achieving the desired open system simulation are reduced
and the capabilities of the same device for executing more complex applications are extended. 
We exemplify our approach of turning a bug into a feature, with the explicitly worked out example of the 
\textit{simulation of the time-evolution of open quantum systems}. We show that it is possible to leverage 
noise processes, intrinsic and added noise, in a current NISQ device when the action of the desired environment 
to be simulated onto the open system can be approximated by a \emph{Markovian stochastic Pauli noise channel}~\cite{breuer2002theory}, but we stress that our methods are not restricted to this specific case.

This paper is organized as follows. In Sec.~II, we discuss the implementation of unitary time-evolution operator on a quantum circuit via the Trotter-
Suzuki product formula and the intrinsic noise of quantum devices. We briefly explain how the intrinsic noise can be transformed to stochastic Pauli noise channels by using Randomized Compiling~\cite{wallman2016noise,hashim2020randomized} and how the corresponding error probabilities can be estimated by Cycle Benchmarking~\cite{erhard2019characterizing} and Error Reconstruction~\cite{flammia2020efficient} techniques. We demonstrate that the Pauli noise channels with estimated error probabilities enable one to simulate open quantum system dynamics under a Lindblad-type noise on NISQ devices in such a way that only the open-system degrees of freedom are encoded in the qubits. We show that the open-system dynamics computed by NISQ devices can be well-described by classical solutions of the Lindblad equation constructed based on the estimated error probabilities of the Pauli noise channels. In Sec.~III, we discuss the conventional Probabilistic Error Cancellation technique~\cite{temme2017error,endo2018practical,sun2021mitigating,van2023probabilistic,cai2022quantum,takagi2022fundamental,suzuki2022quantum,strikis2021learning,guo2022quantum,piveteau2022quasiprobability}, developed to fully mitigate noise in a quantum circuit, and demonstrate that it can also be employed to partially mitigate the error probabilities of stochastic Pauli noise channels in a controlled manner. We show that our approach can be used to implement stochastic Pauli noise with desired decoherence rates on NISQ devices. In Sec.~IV, we show that amplitude damping noise can be implemented efficiently by using reset operations \cite{mi2023stable}, without requiring ancilla qubits and measurement operations. We then generalize our approach to other types of local noise. Lastly, in Sec.~V, we compare our scheme with previous approaches that have been proposed to simulate open quantum systems on quantum devices.

\section{Encoding of the time evolution in a quantum circuit} \label{time_evol_sec}

For simulations of closed system dynamics, various methods have been developed, such as hybrid classical-quantum variational approaches~\cite{gibbs2022dynamical,mizuta2022local,berthusen2022quantum,cirstoiu2020variational}, quantum tensor networks~\cite{niu2022holographic,haghshenas2022variational,foss2021holographic}, and quantum signal processing techniques~\cite{low2019hamiltonian,martyn2021grand}. In this work, we consider the Trotter-Suzuki product formula \cite{childs2021theory,csahinouglu2021hamiltonian} suitable for closed-system simulations on NISQ devices~\cite{clinton2021hamiltonian,childs2018toward}. We demonstrate that it can also be used for efficient quantum simulations of open-system dynamics on a noisy quantum device.

\subsection{Trotter-Suzuki product formula}
The dynamics of a quantum system is governed by a Hamiltonian $\hat{H}$. We consider a Hamiltonian decomposed into $N$ components represented by the tensor products of Pauli matrices $\hat{X},\hat{Y},\hat{Z}$ of multiple qubits
\begin{equation}
    \hat{H} = \sum_{j=1}^{N}\hat{H}_{j},\quad \hat{H}_{j}=\alpha_{j} \hat{P}_{j}, \label{ham_pauli} 
\end{equation}
where $\alpha_{j} \in \mathbb{R}$ and $\hat{P}_{j}=\{\hat{X},\hat{Y},\hat{Z}\}^{\otimes n_j}$ is a Pauli string acting 
on $n_j$ qubits. As an example, we consider the Hamiltonian of a linear chain consisting of $n$ molecules
\begin{align}
    \hat{H} = -&\sum_{m=1}^{n}\frac{E_m}{2}\hat{Z}_m \label{eq:H_PPC}\\
    + &\sum_{m=1}^{n-1}\frac{J_{m,m+1}}{2}(\hat{X}_m \hat{X}_{m+1} + \hat{Y}_m \hat{Y}_{m+1}),
    \nonumber 
\end{align}
where each qubit describes a two-level molecule consisting of ground and excited states, encoded by $\ket{0}$ and $\ket{1}$, respectively, with an energy-gap of $E_m$. The inter-qubit couplings $J_{m,m+1}$ describe a coherent excitation transfer between nearest-neighbour molecules. This model has been widely used to study energy transfer dynamics in various molecular systems, such as photosynthetic complexes and organic solar cells, but a more general form of Hamiltonian can be considered in a noise-assisted digital quantum simulation, including spin and fermionic models, such as the Heisenberg~\cite{guimaraes2022towards,tran2021faster} and Fermi-Hubbard models~\cite{clinton2021hamiltonian,arute2020observation}, respectively.

The product formula describing the time evolution of a quantum system over time $t$ is formally expressed as
\begin{equation}
    e^{-i\hat{H}t} \approx \prod_{d=1}^{D} \hat{U}_{k}(\Delta t), \label{trotter_equation_anyorder}
\end{equation}
where $\hat{U}_{k}(\Delta t)$ represents the unitary time evolution over a finite Trotter time-step $\Delta t$, described by the $k$-th order Trotter-Suzuki product formula, where the first two lowest order examples are given by 
\begin{align}
    \hat{U}_{1}(\Delta t)&=\prod_{j=1}^{N} e^{-i \hat{H}_{j}\Delta t},\label{trotter_equation}\\
    \hat{U}_{2}(\Delta t)&=\left[ \prod_{j=1}^{N} e^{-i \hat{H}_{j}\Delta t/2} \right]\left[\prod_{j'=N}^{1} e^{-i \hat{H}_{j'}\Delta t/2}\right].\label{trotter_equation2}
\end{align}
The total time evolution is then described by $D = t/\Delta t$ Trotter layers where $\hat{U}_{k}(\Delta t)$ is considered in each layer.

An estimate for the number of implemented
Trotter layers $D=t/\Delta t$ required to obtain a Trotter decomposition error $||\hat{U}^{D}_{k}(\Delta t)-e^{-i\hat{H}t}||=O(\varepsilon_{\rm Trot})$ can be given as follows \cite{childs2021theory},
\begin{equation}
    D = O\left(\frac{\alpha_{\rm comm}^{1/k} t^{1+1/k}}{\varepsilon_{\rm Trot}^{1/k}}\right)\label{error_trot}
\end{equation}
for a $k$th-order Trotter-Suzuki product formula, where, $$\alpha_{\rm 
 comm}=\sum_{l_{1},l_{2},\dots,l_{k+1}=1}^{N}||[\hat{H}_{l_{k+1}},\dots [\hat{H}_{l_{2}},\hat{H}_{l_{1}}]\dots]||,$$ and $||\cdot||$ denotes the spectral norm. $\hat{H}_{l_{k}}$ denotes a Pauli string term out of a total of $N$ terms the Hamiltonian is decomposed into (see Eq.~\eqref{ham_pauli}).

We note that variants of Trotter-Suzuki product formulae can also be considered in our approach to reduce simulation error, such as symmetry-protected formulae~\cite{tran2021faster}, random formulae~\cite{chen2021concentration}, and implementation of a specific Trotter sequence of Hamiltonian terms that preserves the locality of the simulated system~\cite{childs2021theory}.

\subsection{Noise characterization}\label{sec: noisy_evol}
On NISQ devices, the implementation of each unitary gate of a Trotterized time-evolution operator, such as $e^{-i\hat{H}_j \Delta t}$ in Eq.~\eqref{trotter_equation}, suffers from noise and dissipation. As a result, the time evolution of a quantum system encoded in the qubits does not only depend on the Hamiltonian implemented on a quantum circuit, but also on the parameters that characterize the intrinsic noise of the quantum device. When its noise can be well-described by a Markovian theory, as demonstrated in previous studies on superconducting quantum computing platforms~\cite{van2023probabilistic,ferracin2022efficiently}, the dynamics of the density matrix $\hat{\rho}(t)$ of the open system encoded in the qubits is described by a Markovian quantum master equation in the form
\begin{equation}
    \frac{d\hat{\rho}(t)}{dt} = \mathcal{L}[\hat{\rho}(t)] = -i[\hat{H},\hat{\rho}(t)]+\mathcal{D}_{\rm intrinsic}[\hat{\rho}(t)], \label{Lindblad_eq}
\end{equation}
where $\mathcal{D}_{\rm intrinsic}[\hat{\rho}(t)]$ represents a Lindblad dissipator describing the intrinsic noise of NISQ devices.

In this work, we aim to implement the decoherence of open quantum systems modelled by a Lindblad equation by harnessing the intrinsic noise of NISQ devices as a resource, rather than encoding environmental degrees of freedom on qubits. Therefore, to simulate open-system dynamics under various decoherence models of interest, we need to characterize the noise channels present in NISQ devices and control the corresponding noise rates. To that end, we apply a noise characterization technique to the noisy implementation $\mathcal{U}$ of the Trotter layer on a quantum circuit (we omitted the $k$-th order and $\Delta t$ dependence of $\mathcal{U}_{k}(\Delta t)$ for simplicity), prior to the implementation of the digital quantum simulation of open-system dynamics. We employ the Randomized Compiling technique~\cite{wallman2016noise,hashim2020randomized} to transform the intrinsic coherent noise in a NISQ device, described by the Kraus operators in the form of $\sum_{j\neq k}\epsilon_{jk}\hat{P}_{j} \hat{\rho} \hat{P}_{k}$, to stochastic Pauli noise channels $\sum_{k}\epsilon_k \hat{P}_{k} \hat{\rho} \hat{P}_{k}$, where $\hat{P}_{k}$ are Pauli strings, including identity operator, and $\epsilon_k$ are the corresponding error probabilities satisfying $\sum_{k}\epsilon_{k}=1$ and $\epsilon_k\ge 0$. Randomized Compiling is implemented on two-qubit gate (CNOT) layers in both noise characterization and digital quantum simulations of open-system dynamics on real IBMQ devices, as explained below.

We employ the Cycle Benchmarking~\cite{erhard2019characterizing}, one of the Randomized Benchmarking methods~\cite{helsen2022general}, together with the Error Reconstruction technique~\cite{flammia2020efficient} to estimate the error probabilities $\epsilon_k$ of quantum circuits on NISQ devices, as recently demonstrated in Ref.~\cite{hashim2020randomized,ferracin2022efficiently,van2023probabilistic}. This enables the characterization of $K$-qubit Pauli noise channels acting on $K$ nearest-neighbour qubits. For instance, when $K=1$, the estimated stochastic Pauli noise channels acting on a single qubit $m$ are expressed as
\begin{equation}
    {\cal E}_{m}^{(1)}(\hat{\rho})=\epsilon_0 \hat{\rho} + \epsilon_X \hat{X}_{m}\hat{\rho}\hat{X}_{m} + \epsilon_Y \hat{Y}_{m}\hat{\rho}\hat{Y}_{m} + \epsilon_Z \hat{Z}_{m}\hat{\rho}\hat{Z}_{m}. \label{eq:E_K1}
\end{equation}
For $K=2$, the stochastic Pauli noise channels acting on two nearest-neighbour qubits $m$ and $m+1$ consist of the single-qubit noise in Eq.~\eqref{eq:E_K1}, independently acting on each qubit, and the correlated two-qubit noise acting on two qubits at the same time
\begin{align}
    {\cal E}_{m,m+1}^{(2)}(\hat{\rho})&={\cal E}_{m}^{(1)}(\hat{\rho})+{\cal E}_{m+1}^{(1)}(\hat{\rho})\\
    &\quad+\epsilon_{XX}\hat{X}_m\hat{X}_{m+1}\hat{\rho}\hat{X}_m\hat{X}_{m+1},\nonumber\\
    &\quad+\epsilon_{XY}\hat{X}_m\hat{Y}_{m+1}\hat{\rho}\hat{X}_m\hat{Y}_{m+1}+\cdots,\nonumber
\end{align}
where the latter takes into account all possible combinations of the Pauli operators of qubits $m$ and $m+1$. Formally, the $K$-qubit noise channels are expressed as
\begin{equation}
    \mathcal{E}^{(K)}(\hat{\rho})=\sum_{k=0}^{4^{K}-1}\epsilon_{k} \hat{P}_{k}\hat{\rho}\hat{P}_{k},\label{stochastic_channel}
\end{equation}
with $\hat{P}_k$ denoting Pauli strings. For a quantum circuit consisting of $n$ qubits, when $K=n$, all possible Pauli strings are considered in the noise characterization. When $K<n$, the $K$-qubit Pauli noise channels can be characterized independently for each subgroup consisting of $K$ nearest-neighbour qubits.

We remark that the Cycle Benchmarking technique can be employed to characterize the noise acting on a quantum circuit $\hat{U}$ satisfying $\hat{U}^{m}=\hat{I}$ for several integer values $m$. The Trotter layer $\hat{U}_{k}(\Delta t)$ constructed based on an open-system Hamiltonian, however, does not satisfy this condition in general. In this work, we modified the parameters of single-qubit gates $\hat{R}_Z(\theta)=e^{-i\theta \hat{Z}/2}$ of the Trotter layer while maintaining its two-qubit gate structure such that $\hat{U}_{k}(\Delta t)$ is a Clifford circuit $\hat{V}$ logically equivalent to the Identity operator. Therefore, the Cycle-Benchmarking condition $\hat{V}^{m}=\hat{I}$ is satisfied for every integer $m$. Since the degree of noise of two-qubit gates is approximately two orders of magnitude higher than that of single-qubit gates on superconducting quantum devices, this approach enables one to estimate the error probabilities $\epsilon_k$ of stochastic Pauli noise acting on the original Trotter layer $\hat{U}_{k}(\Delta t)$ in an accurate manner, as demonstrated below. %\jdg{Recently a no-go theorem for the learnability of stochastic Pauli noise over a Clifford quantum circuit was presented, namely there are always unlearnable degrees of freedom. A recent work \cite{chen2023learnability} presents a no-go thereom for the learnability of stochastic Pauli noise over a Clifford quantum circuit, namely that there are always unlearnable degrees of freedom. We remark that we reduce the transformed Trotter iteration layer to the single operation on which the noise we want to characterize. As mentioned above, the layer is logically equivalent to the Identity operator, hence we do not face the issues of learnability discussed in the abovementioned Ref.~\cite{chen2023learnability}.}

\subsection{Digital quantum simulation of Lindblad model}

\begin{figure*}
\centering
\includegraphics[width=0.9\textwidth]{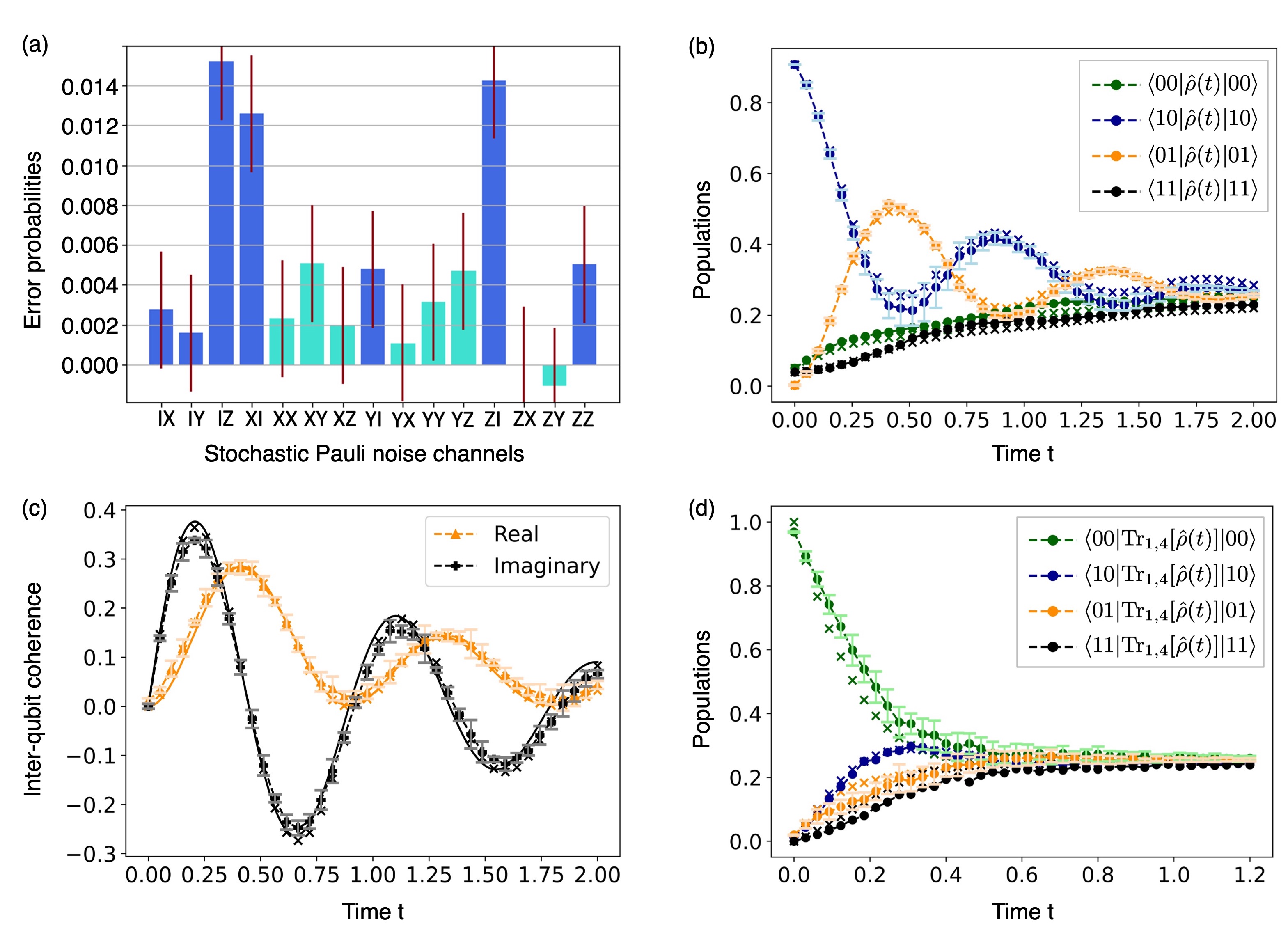}
\caption{(a) Error probabilities $\epsilon_k$ of stochastic Pauli noise channels estimated for a quantum circuit consisting of two qubits ($n=2$) are shown where the real device \textit{ibmq jakarta} with a circuit structure shown in Fig.~\ref{fig:trotter_circuit}(a) was employed. The error probabilities of single-qubit noise channels, $\hat{X}_m$, $\hat{Y}_m$, $\hat{Z}_m$, and two-qubit dephasing $\hat{Z}_{1}\hat{Z}_{2}$ are shown in blue, while the other two-qubit noise channels are displayed in cyan. (b) Population dynamics of qubits computed by the real device \textit{ibmq jakarta} are shown in dots, while those obtained by classically solving a Lindblad equation with decoherence rates determined by the measured error probabilities in (a) are shown in crosses (see the main text). The negative error probabilities found within error bars ($\hat{Z}_{1}\hat{Y}_{2}$) are taken to be zero, and the Lindblad equation was solved by using the first-order Trotter-Suzuki product formula. (c) Real and imaginary parts of inter-qubit coherence dynamics computed by the emulated quantum computer {\it ibmq jakarta} are compared with classical solutions of the Lindblad equation obtained by a standard RK4 solver (see solid lines). (d) For a linear chain consisting of four qubits ($n=4$), the population dynamics of the reduced density matrix ${\rm Tr}_{1,4}[\hat{\rho}(t)]$ of the second and third qubits are displayed, where the results obtained by the real device {\it ibmq lagos} are shown in dots and the classical solutions of the Lindblad equation are shown in crosses. In (a,b) and (d), $R=30$ and $R=22$ randomized compiled circuits were used, respectively (see Appendix~\ref{A}). In all simulations, dimensionless parameters of open-system Hamiltonian in Eq.~\eqref{eq:H_PPC} are taken to be $E_m = 122-0.5 m$ and $J_{m,m+1}=0.5$, motivated by typical electronic parameters of photosynthetic pigment-protein complexes with site energies $12200-50 m\,{\rm cm}^{-1}$ and inter-site electronic coupling strength $50\,{\rm cm}^{-1}$~\cite{mohseni2014quantum}. The initial state is taken to be $\ket{1,0,\cdots,0}$ where the first qubit is in $\ket{1}$, while all the other qubits are in $\ket{0}$.}
\label{fig:evolve_normal}
\end{figure*}

\begin{figure}
%\[\Qcircuit @C=0.8em @R=0.6em{ 
%\text{(a)} \\
%\lstick{\ket{q_{1}}} & \qw & \multigate{1}{e^{-iJ_{12}\hat{X}\hat{X}\Delta t/2}} & \qw & \multigate{1}{e^{-iJ_{12}\hat{Y}\hat{Y}\Delta t/2}} & \qw & \gate{e^{iE_{1}\hat{Z}\Delta t/2}} & \qw\\
%\lstick{\ket{q_{2}}}& \qw & \ghost{e^{-iJ_{12}\hat{X}\hat{X}\Delta t/2}} & \qw & \ghost{e^{-iJ_{12}\hat{Y}\hat{Y}\Delta t/2}} & \qw &\gate{e^{iE_{2}\hat{Z}\Delta t/2}}  & \qw\\
%}\]

%\[\Qcircuit @C=0.8em @R=0.6em{ 
%\text{(b)} \\
%\lstick{\ket{q_{1}}} & \qw & \multigate{1}{e^{-i\hat{H}_{odd}\Delta t}} & \qw & \qw & \qw & \gate{e^{-i\hat{H}_{sing}\Delta t}} & \qw\\
%\lstick{\ket{q_{2}}}& \qw & \ghost{e^{-i\hat{H}_{odd}\Delta t}} & \qw & \multigate{1}{e^{-i\hat{H}_{even}\Delta t}} & \qw & \gate{e^{-i\hat{H}_{sing}\Delta t}} & \qw\\
%\lstick{\ket{q_{3}}}& \qw & \multigate{1}{e^{-i\hat{H}_{odd}\Delta t}} & \qw & \ghost{e^{-i\hat{H}_{even}\Delta t}} & \qw& \gate{e^{-i\hat{H}_{sing}\Delta t}} & \qw\\
%\lstick{\ket{q_{4}}}& \qw &\ghost{e^{-i\hat{H}_{odd}\Delta t}} & \qw & \qw & \qw & \gate{e^{-i\hat{H}_{sing}\Delta t}} & \qw\\
%}\]
\centering
\includegraphics[width=0.48\textwidth]{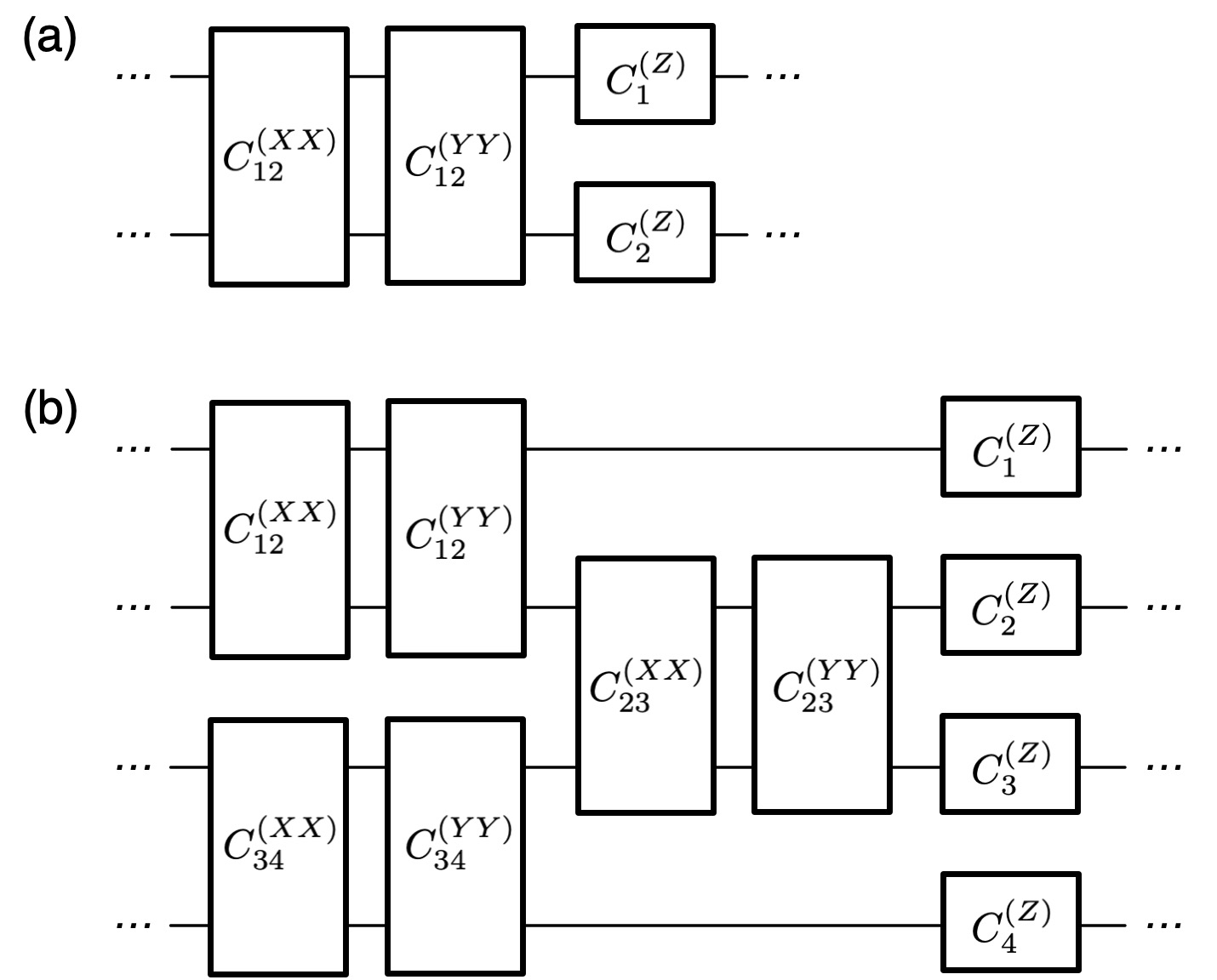}
    \caption{Implementation of a Trotter layer, constructed based on the first-order Trotter-Suzuki product formula together with the Hamiltonian in Eq.~\eqref{eq:H_PPC}, is shown for (a) $n=2$ and (b) $n=4$ qubits. Here $C_{m}^{(Z)}={\rm exp}(iE_m \hat{Z}_{m}\Delta t/2)$, $C_{m,m+1}^{(XX)}={\rm exp}(-iJ_{m,m+1}\hat{X}_{m}\hat{X}_{m+1}\Delta t/2)$ and $C_{m,m+1}^{(YY)}={\rm exp}(-iJ_{m,m+1}\hat{Y}_{m}\hat{Y}_{m+1}\Delta t/2)$ with $\Delta t$ denoting a Trotter time-step.}
    \label{fig:trotter_circuit}
\end{figure}

To demonstrate that the estimated error probabilities $\epsilon_k$ of the stochastic Pauli noise channels can be used to investigate open-system dynamics under a Lindblad-type noise, we consider a Markovian quantum master equation in the form
\begin{equation}
    \frac{d\hat{\rho}(t)}{dt} = \mathcal{L}[\hat{\rho}(t)] = -i[\hat{H},\hat{\rho}(t)]+\mathcal{D}_{\rm stochastic}[\hat{\rho}(t)],\label{eq:L_stochastic_uncontrolled}
\end{equation}
where the Lindblad dissipator $\mathcal{D}_{\rm stochastic}[\hat{\rho}]$ is modelled by
\begin{align}
    \mathcal{D}_{\rm stochastic}[\hat{\rho}] &= \sum_{k=0}^{4^K-1}\gamma_k \left(\hat{P}_k\hat{\rho}\hat{P}_k - \hat{\rho}\right),\label{eq:D_stochastic_uncontrolled}\\
    \gamma_k &= \epsilon_k/\Delta t.\label{eq:gamma_stochastic_uncontrolled}
\end{align}
Here the decoherence rates $\gamma_k$ are defined as a function of the error probabilities $\epsilon_k$ and the Trotter time-step $\Delta t$. 

In the following, we demonstrate by means of an example that the dynamics of a reduced system density matrix $\hat{\rho}_{c}(t)$ simulated by using Eq.~\eqref{eq:L_stochastic_uncontrolled}--\eqref{eq:gamma_stochastic_uncontrolled} on classical computers is well-matched to its counterpart $\hat{\rho}_{q}(t)$ obtained from real and emulated quantum computers, where the system Hamiltonian is implemented by using the first-order Trotter-Suzuki formula in Eq.~\eqref{trotter_equation} and the noise characteristics have been obtained as described above. We note that the Randomized Compiling is applied to every Trotter layer in our approach, so that the stochastic Pauli noise channels, identified by the noise characterization scheme presented in Sec.~\ref{sec: noisy_evol}, are maintained during the digital quantum simulation of open-system dynamics (see Appendix~\ref{A} for more details).

In Fig.~\ref{fig:evolve_normal}(a) and (b), we consider a quantum system encoded in two qubits ($n=2$) modelled by the Hamiltonian in Eq.~\eqref{eq:H_PPC}. The structure of the implemented quantum circuit is shown in Fig.~\ref{fig:trotter_circuit}(a). For a real IBMQ computer, we estimated the error probabilities $\epsilon_k$ of the full stochastic Pauli noise channel ($K=2$). As shown in Fig.~\ref{fig:evolve_normal}(a), we found that the error probabilities $\epsilon_k$ are not uniform with relatively higher values for $\hat{P}_{k}\in\{\hat{X}_{1},\hat{Y}_{1},\hat{Z}_{1},\hat{Z}_{2},\hat{X}_1\hat{Y}_2,\hat{Y}_1\hat{Z}_2,\hat{Z}_1\hat{Z}_2\}$, hinting that qubit 1 is more noisy than qubit 2. As shown in Fig.~\ref{fig:evolve_normal}(b), the population dynamics of the reduced system density matrices $\bra{i,j}\hat{\rho}_{q,c}(t)\ket{i,j}$ simulated by quantum and classical computers are well-matched for all possible values of $i,j\in\{0,1\}$. It is found that when only single-qubit Pauli noise channels are characterized ($K=1$), the quantum and classical results are not matched (not shown here), which can be rationalized based on the fact that the error probabilities of two-qubit Pauli noise channels, such as $\hat{P}_{k}\in\{\hat{X}_1\hat{Y}_2,\hat{Y}_1\hat{Z}_2,\hat{Z}_1\hat{Z}_2\}$, are not negligible. Fig.~\ref{fig:evolve_normal}(c) shows that the inter-qubit coherence dynamics $\bra{10}\hat{\rho}_{q,c}(t)\ket{01}$ are also well-matched, where the emulated noisy quantum device \emph{ibmq jakarta} was used instead of the real IBMQ device due to the long queue waiting time on the IBMQ platform. In Fig.~\ref{fig:evolve_normal}(d), we consider a larger quantum system encoded in four qubits ($n=4$) with a circuit structure shown in Fig.~\ref{fig:trotter_circuit}(b). It is found that the reduced system dynamics $\hat{\rho}_{q}(t)$ computed by the real IBMQ computer are well-matched to the classical solutions $\hat{\rho}_{c}(t)$ of the Lindblad equation when also two-qubit stochastic Pauli noise channels are characterized $(K=2)$. This implies that the Pauli noise channels acting on more than two qubits are negligible in our case due to the short depth of the Trotter layer structure shown in Fig.~\ref{fig:trotter_circuit}(b). We note that the first-order Trotter decomposition is considered in both quantum and classical simulations to investigate the accuracy of the noise characterization scheme considered in our work, independent of the Trotter decomposition error induced by a finite Trotter time-step $\Delta t$. We performed all the simulations with a sufficiently small $\Delta t$, for which classical solutions $\hat{\rho}_{c}(t)$ can also be obtained by using a standard RK4-based solver. These results demonstrate that the intrinsic noise of NISQ devices can be transformed to the stochastic Pauli noise channels via Randomized Compiling, which can be used as a platform to simulate open-system dynamics under a Lindblad-type noise.

\section{Digital control of decoherence in a quantum circuit}\label{control_decoh}

So far we have demonstrated that the intrinsic noise of quantum devices can be transformed to stochastic Pauli noise channels and the corresponding error probabilities $\epsilon_k$ can be estimated, enabling one to implement a Lindblad equation on quantum computers with decoherence rates $\gamma_k=\epsilon_k/\Delta t$ (see Eq.~\eqref{eq:L_stochastic_uncontrolled}--\eqref{eq:gamma_stochastic_uncontrolled}). Now we show that the error probabilities of the Pauli noise channels can be controlled independently by using a partial Probabilistic Error Cancellation (PEC) technique~\cite{endo2018practical,sun2021mitigating}, leading to reduced error probabilities $\epsilon_k (1-r_k)$ with independent partial mitigation factors $r_k \in [0,1]$. This approach makes it possible to implement an arbitrary set of target decoherence rates $\Gamma_k$ on NISQ devices, so that one can investigate open-system dynamics described by a Lindblad equation of interest in the following form
\begin{align}
    \mathcal{D}_{\rm stochastic}^{\rm (controlled)}[\hat{\rho}] &= \sum_{k=0}^{4^K-1}\Gamma_k \left(\hat{P}_k\hat{\rho}\hat{P}_k - \hat{\rho}\right),\label{eq:D_stochastic_controlled}\\
    \Gamma_k &=\epsilon_k (1-r_k)/\Delta t.\label{eq:gamma_stochastic_controlled}
\end{align}
We note that our method can be modified to implement more general Lindblad noise models beyond the stochastic Pauli noise channels in Eq.~\eqref{eq:D_stochastic_controlled}, such as amplitude damping noise as will be discussed later.

\subsection{Probabilistic Error Cancellation}
Here we provide a brief overview of PEC technique~\cite{temme2017error,endo2018practical,sun2021mitigating,van2023probabilistic,cai2022quantum,takagi2022fundamental,suzuki2022quantum,strikis2021learning,guo2022quantum,piveteau2022quasiprobability}. For more details, we refer the readers to recent reviews such as Refs.~\cite{temme2017error,endo2018practical,cai2022quantum}.
PEC starts by identifying the noise channel $\mathcal{E}$ acting on 
an ideal, noiseless circuit $\mathcal{C}(\hat{\rho})=\hat{U}_{k}(\Delta t)\hat{\rho}\hat{U}_{k}^{\dagger}(\Delta t)$, for instance, describing the Hamiltonian dynamics of a quantum system over a Trotter time-step $\Delta t$. As discussed in Sec.~\ref{sec: noisy_evol}, one can characterize the noise of a quantum circuit, yielding $K$-qubit stochastic Pauli noise channels, $\mathcal{E}(\hat{\rho})=\sum_{k=0}^{4^{K}-1}\epsilon_{k}\mathcal{P}_{k}(\hat{\rho})=\sum_{k=0}^{4^{K}-1}\epsilon_{k} \hat{P}_{k}\hat{\rho} \hat{P}_{k}$ with $\hat{P}_{0}=\hat{I}^{\otimes K}$ denoting identity operator. To fully mitigate the noise, the conventional PEC has considered the inverted noise channel $\mathcal{E}^{-1}$ acting on the noisy quantum circuit, namely $\mathcal{E}^{-1}$ applied to $\mathcal{U}(\hat{\rho}) = \mathcal{E}\cdot\mathcal{C}(\hat{\rho})$. Since $\mathcal{E}^{-1}$ is not a complete-positive (CP) map, one cannot physically implement it, hence one cannot perfectly cancel in practice the stochastic Pauli noise channels in $\mathcal{U}$. %However, when the error probabilities $\epsilon_k$ are sufficiently low, the conventional PEC technique can almost fully mitigate the noise (up to first order on $\epsilon = \sum_{k>0}\epsilon_{k}$) as $\mathcal{C}(\hat{\rho})\approx \mathcal{E}^{-1}\mathcal{U}(\hat{\rho})$ by implementing the non-physical inverted noise channel in a probabilistic manner.

\begin{figure}
\centering
\includegraphics[width=0.48\textwidth]{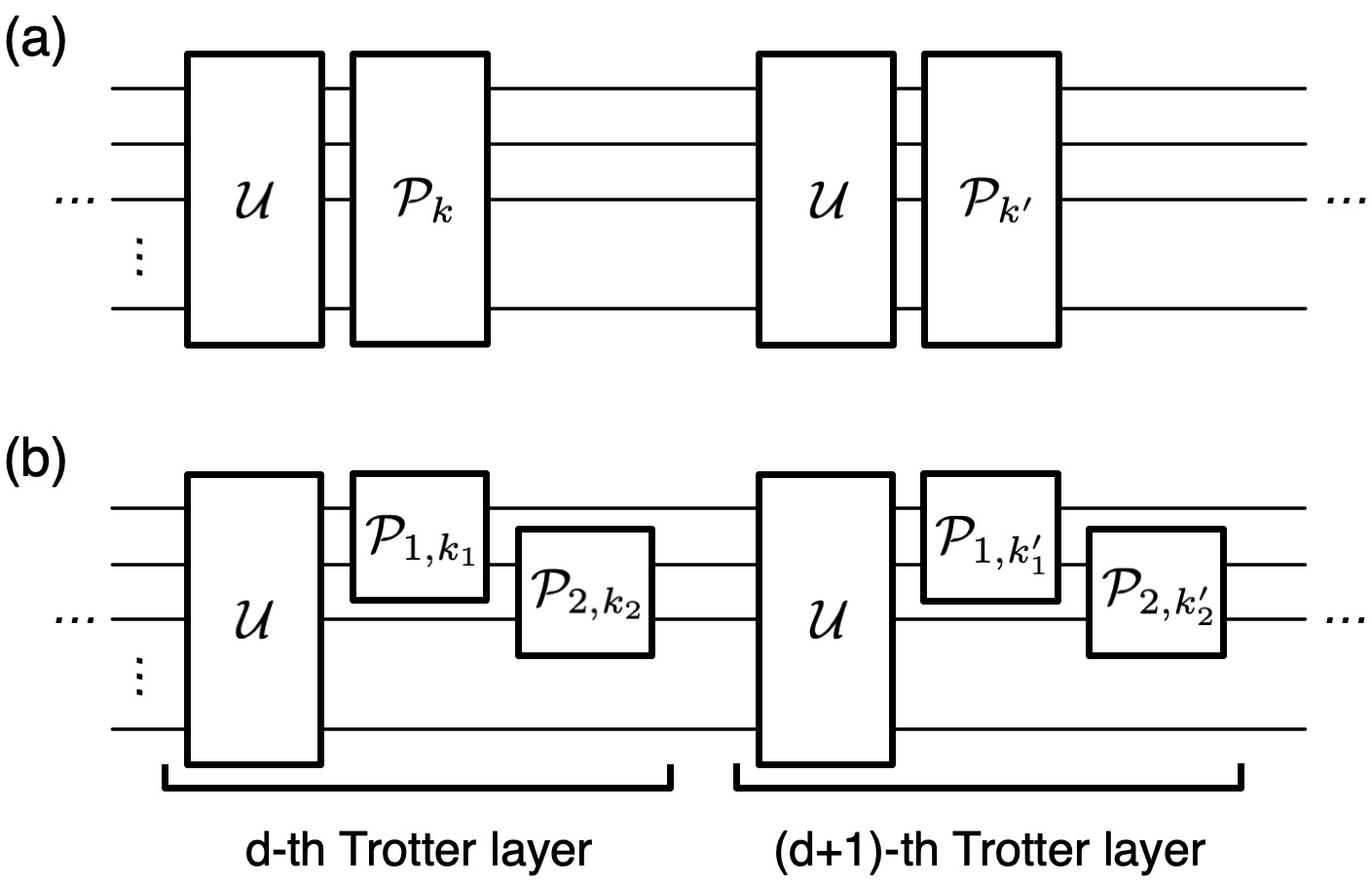}
\caption{Probabilistic Error Cancellation technique applied to a quantum circuit consisting of $n$ qubits and multiple Trotter layers. (a) To mitigate $K$-qubit Pauli noise channels with $K=n$, one of the Pauli strings $\hat{P}_{k}$ is randomly generated based on a probability distribution $p_{k}^{\rm (PEC)}$ for each Trotter layer independently. (b) When $K=2$, one of the Pauli strings $\hat{P}_{m,k_m}$ acting on only one or two qubits is randomly generated for every pair of nearest-neighbour qubits $m$ and $m+1$.}
\label{fig:PEC}
\end{figure}

When $K$-qubit Pauli noise channels act on a quantum circuit consisting of $K$ qubits ($n=K$), the inverted noise channel is written as,
\begin{align}  \mathcal{E}^{-1}=\sum_{k=0}^{4^{K}-1}q_{k}\mathcal{P}_{k}=C_{\rm mit}\sum_{k=0}^{4^{K}-1}p^{\rm (PEC)}_{k}\text{sign}(q_{k})\mathcal{P}_{k}, \label{eq:PEC}
\end{align}
where $q_{0}=1+\sum_{k>0}\epsilon_{k}$, $q_{k>0}=-\epsilon_{k}$, $p_{k}^{\rm (PEC)} = |q_{k}|/C_{\rm mit}$ with $C_{\rm mit}=\sum_{k=0}|q_{k}|$ called the mitigation cost. In the probabilistic application of the non-CP map $\mathcal{E}^{-1}$, one of the Pauli operators $\hat{P}_{k}$ is randomly chosen based on the probabilities $p_{k}^{\rm (PEC)}$ and then applied to a Trotter layer. For a quantum circuit with $D=t/\Delta t$ Trotter layers, the probabilistic non-CP map is applied $D$ times with the Pauli operators $\hat{P}_{k}$ independently sampled for each Trotter layer, as schematically shown in Fig.~\ref{fig:PEC}(a).

PEC can also be applied to a quantum circuit consisting of $n$ qubits where $K$-qubit Pauli noise channels with $K<n$ act on several subgroups consisting of $K$ qubits. Here the noise mitigation can be implemented by considering multiple inverted noise channels acting on different subgroups. Fig.~\ref{fig:PEC}(b) shows an example of $n>K=2$ where the inverted noise channels act on every pair of nearest-neighbour qubits. We note that when different $K$-qubit Pauli noise channels act on the qubits in the intersection of different subgroups, one needs to adjust the mitigation probabilities $p_{k}^{\rm (PEC)}$ in such a way that the noise acting on the shared qubits is not cancelled multiple times. 

In PEC, the outcome of an observable $\hat{O}$ measured on a noise-mitigated quantum circuit is multiplied by a product of mitigation costs and other prefactors in Eq.~\eqref{eq:PEC}, $\prod_{d=1}^{D}\prod_{m}C^{(m)}_{\rm mit}\text{sign}(q_{k}^{(m,d)})$, where $m$ describes different subgroups of qubits under the action of $K$-qubit Pauli noise channels. The value of $\text{sign}(q_{k}^{(m,d)})$ depends on which Pauli operator $\hat{P}_k$ is randomly sampled in the $d$-th Trotter layer. The total mitigation cost of quantum simulation is defined as
\begin{equation}
    C_{\rm tot}=\prod_{d=1}^{D}\prod_{m}C^{(m)}_{\rm mit}.\label{cost_definition}
\end{equation}
For a noise-mitigated density matrix $\hat{\rho}_{\rm mit}(t)$, the expectation value $\text{Tr}[\hat{O}\hat{\rho}_{\rm mit}(t)]$ of the observable $\hat{O}$ is obtained by classically averaging the outcomes of the PEC scheme, requiring multiple copies of quantum circuits.

\subsection{Decoherence rate control scheme}\label{sec:decoherence_control}

So far we have discussed the conventional PEC that aims to fully mitigate the stochastic Pauli noise channels~\cite{endo2018practical,sun2021mitigating}. Contrary to the previous studies on PEC, here we aim to partially cancel the noise, so that the controlled error probabilities of the Pauli noise channels can be used as a resource for open-system simulations. To that end, we consider $q_{0}=1+\sum_{k>0}r_{k}\epsilon_{k}$ and $q_{k>0}=-r_{k}\epsilon_{k}$ with $r_k \in [0,1]$, renormalizing the probabilities of the PEC scheme, namely $p_{k}^{\rm (PEC)} = |q_{k}|/C_{\rm mit}$ with $C_{\rm mit}=\sum_{k=0}|q_{k}|$. The corresponding {\it partially} inverted noise channel enables one to reduce the Pauli error probabilities in a controlled manner, $\epsilon_k\rightarrow (1-r_k)\epsilon_k$. The reduced error probabilities make it possible to implement the Lindblad equation in Eq.~\eqref{eq:D_stochastic_controlled} on NISQ devices with controlled decoherence rates $\Gamma_k=\epsilon_k(1-r_k)/\Delta t$.

\begin{figure}
\centering
\includegraphics[width=0.48\textwidth]{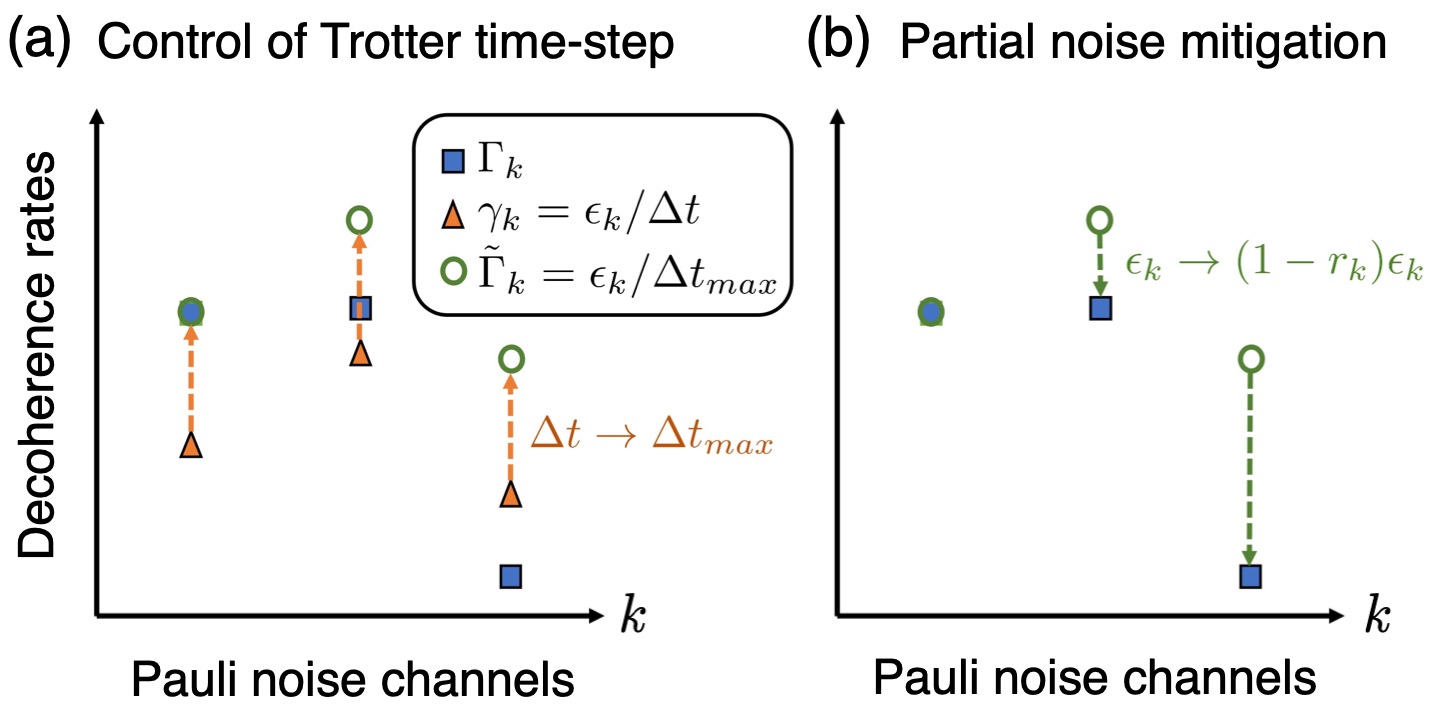}
\caption{Decoherence rate control scheme consisting of two steps. (a) For a given set of Pauli error probabilities $\epsilon_k$, a Trotter time-step $\Delta t$ is decreased so that $\gamma_k=\epsilon_k/\Delta t$ becomes larger than or equal to target decoherence rates $\Gamma_k$. The maximum value of the Trotter time-step satisfying this condition is denoted by $\Delta t_{\rm max}$. (b) If $\tilde{\Gamma}_{k}=\epsilon_k/\Delta t_{\rm max}$ is larger than the target decoherence rate $\Gamma_{k}$ for some $k$, the corresponding error probability $\epsilon_k$ is partially mitigated by using Probabilistic Error Cancellation, so that the target decoherence rate $\Gamma_{k}=\epsilon_k (1-r_k)/\Delta t_{\rm max}$ is implemented with an optimal mitigation factor $r_k$.}
\label{fig:mitigation_diagram}
\end{figure}

For a given set of target decoherence rates $\Gamma_{k}$, our scheme works as follows. Before our PEC scheme is applied to a quantum circuit, for a given Trotter time-step $\Delta t$, the decoherence rates $\gamma_k=\epsilon_k/\Delta t$ of the stochastic Pauli noise channels may satisfy $\Gamma_{k}>\gamma_k$ for some $k$. Since the error probabilities $\epsilon_k$ of the Pauli noise channels can be decreased but not increased by our PEC scheme, namely $\epsilon_k\rightarrow (1-r_k)\epsilon_k$ with $r_k \in [0,1]$, the decoherence rates $\gamma_k$ implemented on the quantum device can be increased only by reducing the Trotter time-step $\Delta t$. Hence we decrease $\Delta t$ until $\Gamma_{k}\le \gamma_{k}=\epsilon_{k}/\Delta t$ is satisfied for all $k$ (see Fig.~\ref{fig:mitigation_diagram}(a)). Here one can find a range of $\Delta t \in (0,\Delta t_{\rm max}]$ satisfying this condition and take the maximum value $\Delta t_{\rm max}$ to minimize the number $D=t/\Delta t_{\rm max}$ of Trotter layers. After that, for some $k$ where the target decoherence rates are lower than the implemented decoherence rates, namely $\Gamma_{k}<\tilde{\Gamma}_{k}=\epsilon_k/\Delta t_{\rm max}$, we apply our PEC scheme to partially mitigate the corresponding error probabilities $\epsilon_{k}$, so that $\Gamma_{k}=\epsilon_k (1-r_k)/\Delta t_{\rm max}$ holds for all $k$ (see Fig.~\ref{fig:mitigation_diagram}(b)). In this way, one can implement arbitrary target decoherence rates $\Gamma_k$, in principle, on NISQ devices.

We note that the error probabilities $\epsilon_k$ of real quantum devices are not uniform, as shown in Fig.~\ref{fig:evolve_normal}(a). This implies that even a simple decoherence model with uniform target decoherence rates $\Gamma_k$ requires our {\it noise-specific} mitigation scheme with non-uniform values of $r_k$. In addition, the control of the Trotter time-step $\Delta t$ alone is not sufficient to implement arbitrary target decoherence rates, especially when $\Gamma_k$ are small, as a larger $\Delta t$ results in a higher Trotter decomposition error. This is indeed the main issue of Ref.~\cite{leppakangas2022quantum}.

\subsection{Implementation} \label{sec: implementation}

\begin{figure*}
\centering
\includegraphics[width=0.9\textwidth]{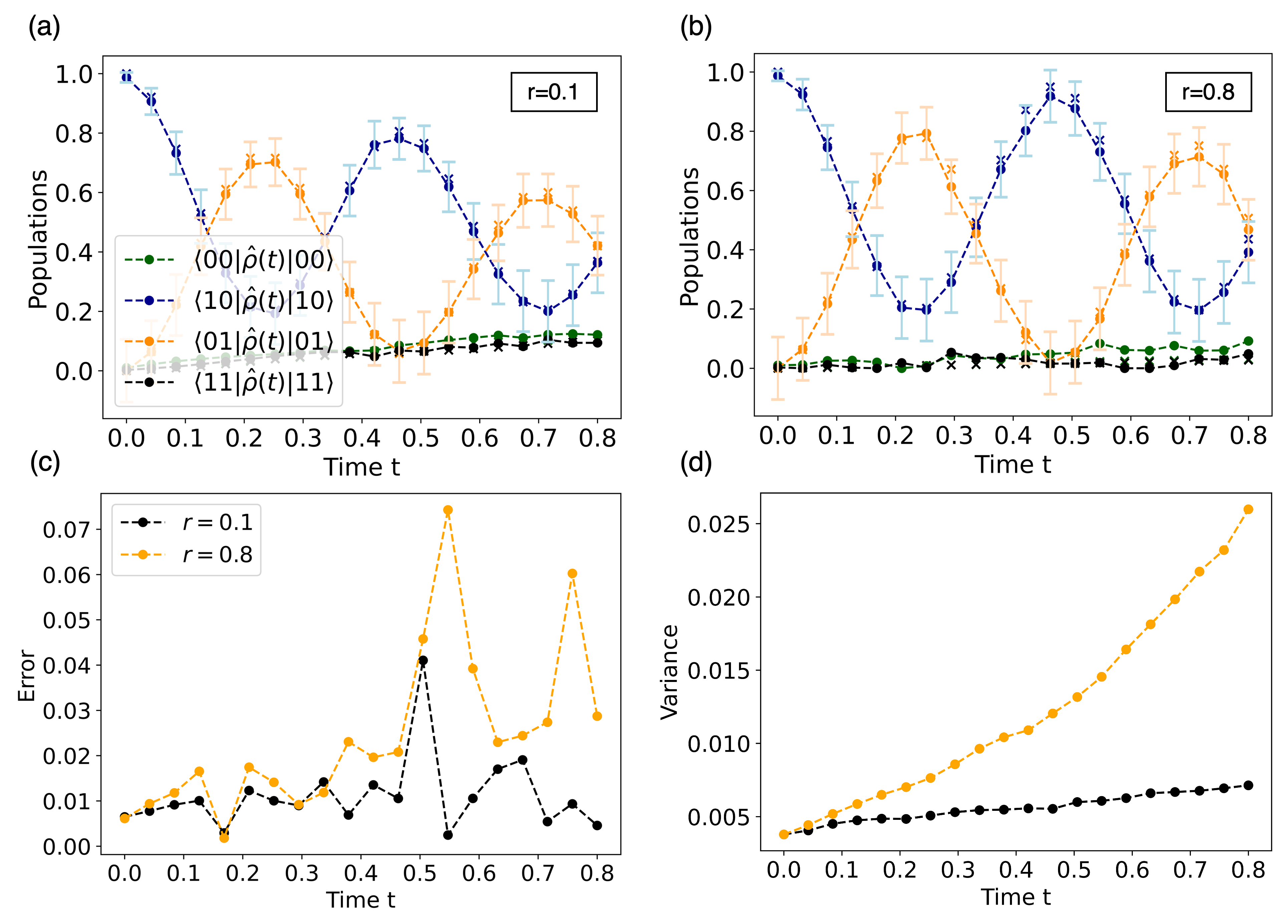}
\caption{(a,b) Population dynamics of an open quantum system encoded in two qubits ($n=2$), computed by the emulated noisy IBMQ device with partial noise mitigation scheme. In (a) and (b), uniform mitigation factor $r=0.1$ and $r=0.8$ were considered, respectively, where $90 C_{\rm tot}^{2}$ circuits were employed to partially mitigate noise with $C_{\rm tot}^{2}$ computed by using Eq.~\eqref{Ctot_simple}. As shown in Eq.~\eqref{cost_scaling2}, the total mitigation factor $C_{\rm tot}$ increases exponentially as a function of the number $D$ of Trotter layers and the uniform mitigation factor $r$. This implies that the partial PEC cost increases as a function of time $t$, and the number $90 C_{\rm tot}^{2}$ of circuits considered in simulations is smaller for $r=0.1$ than for $r=0.8$. The simulated results obtained by the emulated noisy IBMQ device are well matched to classical solutions of the Lindblad equation in Eq.~\eqref{eq:D_stochastic_controlled} with controlled decoherence rates in Eq.~\eqref{eq:gamma_stochastic_controlled}, shown in crosses. (c,d) To demonstrate that the number of circuits required for the partial PEC scheme depends on the uniform mitigation factor $r$, the same number of circuits was considered for $r=0.1$ and $r=0.8$ in independent simulations, specifically $50$ circuits. To quantify the error introduced by a finite number of circuits considered in the partial PEC scheme, the difference in population dynamics simulated by quantum and classical computers is considered (see the main text), leading to (c) average error and (d) its standard deviation. The Hamiltonian parameters considered in simulations are as in Fig.~\ref{fig:evolve_normal}.}
\label{fig:error_implementation}
\end{figure*}

To demonstrate that our partial error mitigation scheme can be used to implement the Lindblad equation in Eq.~\eqref{eq:D_stochastic_controlled} on NISQ devices with controlled decoherence rates in Eq.~\eqref{eq:gamma_stochastic_controlled}, here we consider a quantum system encoded in two qubits ($n=2$). For simplicity, we consider a uniform mitigation factor $r=r_k$ for all $k$. We used the emulated noisy IBMQ device \emph{ibmq lagos} as our testbed.

For two different mitigation factors $r=0.1$ and $r=0.8$, respectively, Fig.~\ref{fig:error_implementation}(a) and (b) show the population dynamics $\bra{i,j}\hat{\rho}_{q}(t)\ket{i,j}$ with $i,j\in\{0,1\}$ simulated by the emulated quantum computer, which are well-matched to the solutions $\hat{\rho}_{c}(t)$ of the Lindblad equation solved on classical computers via a first-order Trotter-Suzuki product formula so that the Trotter decomposition error is identical in quantum and classical simulations (coherence dynamics are also well-matched but not shown here). As expected, the open-system dynamics becomes more coherent with a slower decay of oscillations for a larger mitigation factor $r$. Here the number of samples considered in our generalized PEC scheme is increased until the quantum and classical results are well-matched, and it is found that a larger number of samples is required for a higher mitigation factor $r$.

To demonstrate the dependence of the sampling cost of our PEC scheme on the mitigation factor $r$, Fig.~\ref{fig:error_implementation}(c) shows the difference in population dynamics simulated by quantum and classical computers, quantified by
\begin{equation}
  \eta(\hat{\rho}_{q}(t),\hat{\rho}_{c}(t)) = \frac{1}{4}\sum_{i=0}^{1}\sum_{j=0}^{1}|\bra{i,j}\hat{\rho}_{q}(t)-\hat{\rho}_{c}(t)\ket{i,j}|,\label{error_eq_impl}
\end{equation}
for the mitigation factors $r=0.1$ and $r=0.8$. Here the number of samples of our PEC scheme is taken to be independent of the mitigation factor $r$, and the classical solutions $\hat{\rho}_{c}(t)$ of the Lindblad equation were obtained by using the first-order Trotter-Suzuki product formula. It is notable that the average difference $\eta(\hat{\rho}_{q}(t),\hat{\rho}_{c}(t))$ between quantum and classical results is larger for higher $r$. In addition, as shown in Fig.~\ref{fig:error_implementation}(d), the variance of $\eta(\hat{\rho}_{q}(t),\hat{\rho}_{c}(t))$ is also larger for higher $r$, hinting that the sampling cost of our PEC scheme, required to obtain reliable open-system dynamics, increases as a function of the mitigation factor $r$, as will be analyzed later in more detail.

These results demonstrate that a Lindblad equation with controlled decoherence rates can be implemented on NISQ devices by our partial error mitigation scheme. We note that quantum and classical results are also well-matched when a larger quantum system is considered, encoded in four qubits ($n=4$), together with non-uniform mitigation factors $r_k$, as shown in Appendix~\ref{D}.

\subsection{Resource scaling of partial noise mitigation}\label{scalingPEC}

The application of PEC to large quantum circuits faces a bottleneck due to the enhanced statistical fluctuations. For a given observable $\hat{O}$, when a finite number $M$ of outcomes is measured in experiments without PEC, the uncertainty in expectation value $\langle \hat{O}\rangle_M$ may be described by its variance $\Delta \hat{O}_M\propto M^{-1}$. It has been shown that when PEC is employed, the variance is approximately increased to $\Delta \hat{O}_{M}^{\rm (PEC)}\propto C_{\rm tot}^{2} M^{-1}$ where $C_{\rm tot}\ge 1$ is the total mitigation cost in Eq.~\eqref{cost_definition}~\cite{endo2018practical,sun2021efficient}. Therefore, to maintain the degree of the uncertainty in expectation values, one needs to increase the number of single-shot measurements from $M$ to $M C_{\rm tot}^{2}$ when PEC is employed. For the full mitigation scheme with $r_k=1$ for all $k$, it can be shown that the total mitigation cost $C_{\rm tot}$ increases exponentially as a function of the number $n$ of qubits, the depth of quantum circuits, and the total error probability $\sum_{k>0}\epsilon_k$ of stochastic Pauli noise channels.

\begin{figure*}
\centering
\includegraphics[width=0.9\textwidth]{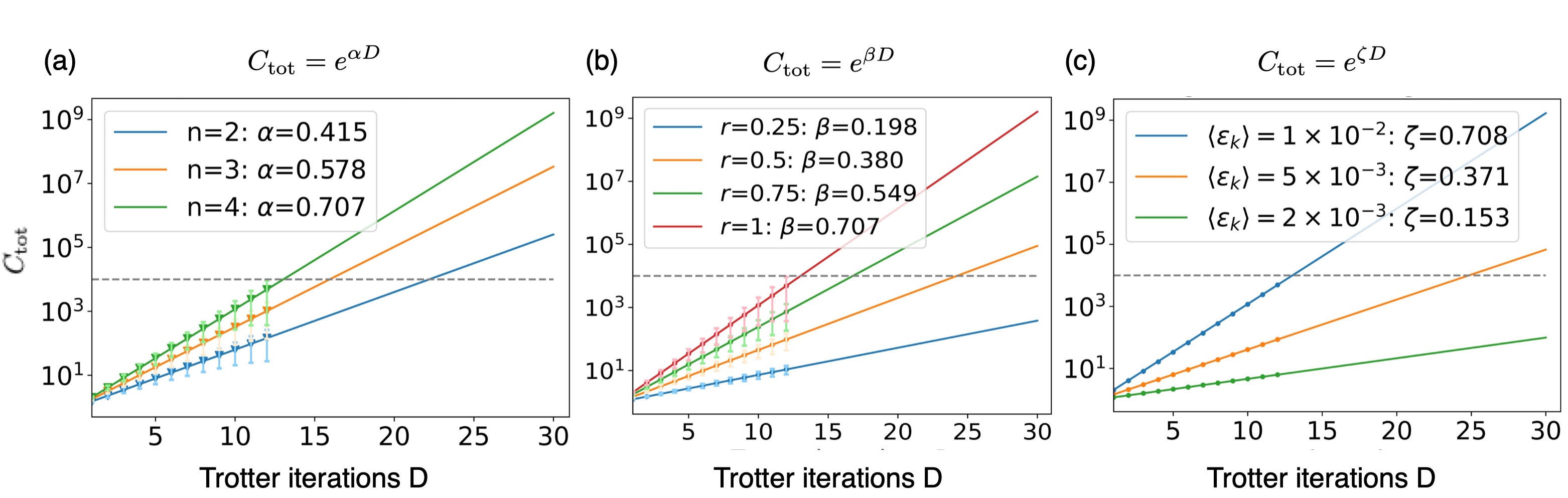}
\caption{(a,b) For the Pauli error probabilities estimated from real device \textit{ibmq lagos}, (a) the total mitigation cost $C_{\rm tot}$ is shown as a function of the number $D$ of Trotter layers for different numbers of qubits, $n\in\{2,3,4\}$, with a fixed uniform mitigation factor $r=1$. The total mitigation cost $C_{\rm tot}$ is well fitted by an exponential function $e^{\alpha D}$, as shown in solid lines, with the values of $\alpha$ shown in the inset. The dependence of $C_{\rm tot}$ on the number $n$ of qubits can be well described by $\alpha \approx (0.186\pm 0.007)n$. (b) $C_{\rm tot}$ is shown as a function of $D$ for several values of uniform mitigation factor, $r\in\{0.25,0.50,0.75,1.00\}$, where the number of qubits is taken to be $n=4$. $C_{\rm tot}$ can be well fitted by an exponential function $e^{\beta D}$, and the dependence of $C_{\rm tot}$ on the uniform mitigation factor $r$ is well described by $\beta\approx (0.678\pm 0.013)r$. (c) $C_{\rm tot}$ is shown as a function of $D$ for $n=4$ and $r=1$ where the Pauli error probabilities were randomly generated from Gaussian distributions with equal average $\langle\epsilon_k \rangle$ and FWHM $\Delta\epsilon_k = \frac{1}{2}\langle \epsilon_k \rangle$ for all $k$. Horizontal dashed lines indicate the maximum cost supported by current quantum devices, defined by the number of circuits that can be executed in $24\,$h~\cite{bravyi2022future}.}
\label{fig:scaling_res}
\end{figure*}

For the case that the error probabilities $\epsilon_k$ are partially reduced with the mitigation factors $r_k \in [0,1]$, the total mitigation cost $C_{\rm tot}$ in Eq.~\eqref{cost_definition} can be computed analytically
\begin{align}
    C_{\rm tot} = \prod_{d=1}^{D}\prod_{m}(1+2\epsilon_r)= (1+2\epsilon_r)^{g(n) D},\label{Ctot_simple}
\end{align}
where $\epsilon_{r}=\sum_{k>0}\epsilon_{k}r_{k}$ is the sum of the error probabilities $\epsilon_k$ weighted by the corresponding mitigation factors $r_k$, and $g(n)$ is the number of $K$-qubit Pauli noise channels acting on a quantum circuit consisting of $n$ qubits. For simplicity, we assume that $\epsilon_r$ are identical for all the $K$-qubit stochastic Pauli noise channels acting on different subgroups of qubits ($n>K$). Practically we are interested in the limit of $n\gg K$ where a large open-system is encoded in a quantum circuit, and the noise is not strongly correlated amongst many qubits and therefore well-characterized by small $K\approx 2$. In this case, the number of different $K$-qubit Pauli noise channels acting on nearest-neighbour qubits increases linearly as a function of the total number $n$ of qubits, $g(n)\propto n$. In the limit of a small total error probability $\epsilon_r$ weighted by the mitigation factors and a sufficiently large number of qubits and/or Trotter layers, namely $\epsilon_r\rightarrow 0$ and $nD \rightarrow \infty$, the total mitigation cost can be approximately described by an exponential function $C_{\rm tot}\sim e^{2\epsilon_r n D}$. For the finite values of $\epsilon_r$, $n$ and $D$, now we show that the total mitigation factor can also be well-described by an exponential function in the form
\begin{equation}
    C_{\rm tot} \sim e^{\lambda n D \epsilon_{r}}, \label{cost_scaling2}
\end{equation}
with a positive constant $\lambda$ introduced as a fitting parameter of simulated results.

In Fig.~\ref{fig:scaling_res}(a) and (b), we show the total mitigation cost $C_{\rm tot}$ in a logarithmic scale, computed based on its definition in Eq.~\eqref{cost_definition}, as a function of the number $D$ of Trotter layers. The error probabilities $\epsilon_k$, determining the cost function $C_{\rm tot}$, were obtained from a real IBMQ computer (see Sec.~\ref{sec: noisy_evol}). For simplicity, we consider a uniform mitigation factor $r$, satisfying $r=r_k$ for all $k$. The maximum cost supported by current quantum devices is highlighted by horizontal dashed lines. In Fig.~\ref{fig:scaling_res}(a), where $r=1$, $C_{\rm tot}$ is displayed for different sizes of quantum circuits, $n\in \{2,3,4\}$. It is found that the numerical results can be well fitted by an exponential function in the form $e^{\alpha(n) D}$ with $\alpha(n) \approx 0.186\,n$, demonstrating the linear dependence of $\log(C_{\rm tot})$ on the number $n$ of qubits. In Fig.~\ref{fig:scaling_res}(b), where $n=4$, $C_{\rm tot}$ is shown for different uniform mitigation factors $r\in \{0.25,0.50,0.75,1.00\}$, where numerical results can be well-fitted by $e^{\beta(r) D}$ with $\beta\approx 0.678\,r$, revealing the linear dependence of $\log(C_{\rm tot})$ on the partial mitigation factor $r$. These results are in line with the approximate form of the total mitigation cost $C_{\rm tot}$ in Eq.~\eqref{cost_scaling2}.

We note that the total mitigation cost $C_{\rm tot}$ of our generalized PEC scheme decreases exponentially, as the partial mitigation factor $r<1$ is reduced, when compared to the conventional full mitigation scheme with $r=1$. This implies that our PEC scheme can be applied to a larger quantum circuit consisting of an increased number of qubits and/or Trotter layers, hinting that digital quantum simulation of open-system dynamics via our technique may be a promising application to NISQ devices. It is also notable that the total mitigation cost can be exponentially decreased further as the noise probabilities $\epsilon_k$ are reduced in the future quantum devices. To highlight this aspect, in Fig.~\ref{fig:scaling_res}(c), where $n=4$, we computed $C_{\rm tot}$ based on randomly generated error probabilities $\epsilon_k$ from Gaussian distributions with equal average $\langle\epsilon_k\rangle$ and FWHM $\Delta\epsilon_k=\frac{1}{2}\langle\epsilon_k\rangle$ for all $k$. It is found that when $\langle\epsilon_k\rangle=0.01$, this model can quantitatively reproduce the total mitigation cost computed based on the error probabilities of a real IBMQ computer (see the case of $n=4$ in Fig.~\ref{fig:scaling_res}(a)). As the average error probabilities $\langle\epsilon_k\rangle$ are reduced from 0.01, via 0.005, to 0.002, the total mitigation cost decreases exponentially, as shown in Fig.~\ref{fig:scaling_res}(c).

These results demonstrate that our partial noise mitigation scheme can be employed to implement a Lindblad model with arbitrary target decoherence rates $\Gamma_k$ on NISQ devices, but it requires a sufficiently large number of samples that scales with $C_{\rm tot}^{2}$. In real quantum devices, however, the maximum number $M_{\rm max}^{\rm (NISQ)}$ of circuits that can be executed within a given period of time is finite. According to Ref.~\cite{bravyi2022future}, $M_{\rm max}^{\rm (NISQ)}\sim 10^{8}$ for moderate circuit depths within a day. Therefore, the number of samples required for our technique should satisfy $C_{\rm tot}^{2}\sim e^{\lambda n D \epsilon_r}\lesssim M_{\rm max}^{\rm (NISQ)}$. For a given set of target decoherence rates $\Gamma_k$, this inequality can be expressed as
\begin{equation}
    \epsilon \lesssim \frac{\Gamma t}{D} + \frac{\ln(M_{\rm max}^{\rm (NISQ)})}{2\lambda n D^2}, \label{scaling_ctot}
\end{equation}
where $\epsilon=\sum_{k>0}\epsilon_k$ denotes the total Pauli noise probability, $\Gamma=\sum_{k>0}\Gamma_{k}$ the total target decoherence rate, and $t$ the simulation time with a Trotter time-step $\Delta t\le \Delta t_{\rm max}$ (see Sec.~\ref{sec:decoherence_control}). The first term in Eq.~\eqref{scaling_ctot} shows that the total Pauli noise probability $\epsilon$ scales linearly with the total target decoherence rate $\Gamma$. On one hand, this implies that when the total target decoherence rate $\Gamma$ is sufficiently high, the corresponding Lindblad model can be readily implemented on a quantum device even if its total noise probability $\epsilon$ is high. On the other hand, when the total target decoherence rate $\Gamma$ is low, the quantum device should have a sufficiently low total noise probability $\epsilon$, or the Pauli noise channels should be heavily mitigated, requiring a high sampling cost. The second term in Eq.~\eqref{scaling_ctot} shows that the total Pauli noise probability $\epsilon$ allowed by our partial noise mitigation scheme may be linearly increased by exponentially enhancing the capability of a quantum device to run several copies of circuits within a given period of time, quantified by $M_{\rm max}^{\rm (NISQ)}$. Note that the second term in Eq.~\eqref{scaling_ctot} is inversely proportional to the number $n$ of qubits and the square of the number of Trotter layers, $D^2$, implying that a higher $M_{\rm max}^{\rm (NISQ)}$ is required for digital quantum simulations of a larger open quantum system on a longer time scale.

We remark that the number $D$ of Trotter layers, required to achieve a desired Trotter decomposition error $\varepsilon_{\rm Trot}$ in Eq.~\eqref{error_trot}, depends on the structure of the Hamiltonian of a target open quantum system, the simulation time $t$, and the order $k$ of the Trotter-Suzuki product formula considered in simulations. As an example, we may consider the first order product formula ($k=1$), and a linear chain (or square grid) structure of qubits with a uniform nearest-neighbor coupling strength $J$ and on-site energy $E$ (see Eq.~\eqref{eq:H_PPC}). In this case, the number $D$ of Trotter layers can be expressed as
\begin{equation}
    D = O\left(\frac{n^d J (J+E) t^{2}}{\varepsilon_{\rm Trot}}\right),\label{scaling_trot}
\end{equation}
where $d=1$ (or 2) for the linear chain (or two-dimensional square grid) structure (for other classes of Hamiltonian, see Ref.~\cite{childs2021theory}). Using Eq.~\eqref{scaling_trot}, now we can express Eq.~\eqref{scaling_ctot} as a function of the Trotter decomposition error $\varepsilon_{\rm Trot}$
\begin{equation}
    \epsilon \lesssim O\left(\frac{\varepsilon_{\rm Trot}\Gamma}{n^d J(J+E)t} + \frac{\varepsilon_{\rm Trot}^{2}\ln(M_{\rm max}^{\rm (NISQ)})}{2\lambda n^{2d+1}J^2 (J+E)^2 t^4}\right).
\end{equation}
Note that the total Pauli noise probability $\epsilon$ allowed by our scheme increases as a function of the Trotter decomposition error $\varepsilon_{\rm Trot}$, and it is inversely proportional to the parameters $J$ and $E$ of the open-system Hamiltonian, and the simulation time $t$.

\section{Amplitude damping}
So far we have demonstrated that the stochastic Pauli noise models in Eq.~\eqref{eq:D_stochastic_controlled} can be implemented by using our partial noise mitigation scheme. Here we show how the Lindblad models beyond the stochastic Pauli noise can be implemented on quantum devices with controlled decoherence rates, such as amplitude damping that has been widely considered in classical simulations of open-system dynamics~\cite{breuer2002theory}. We demonstrate that local noise models can be efficiently implemented by using reset operations without introducing costly ancilla qubits.

\subsection{Amplitude damping noise with ancilla qubits}\label{sec:amplitude_damping_with_ancilla}

\begin{figure}
\centering
\includegraphics[width=0.48\textwidth]{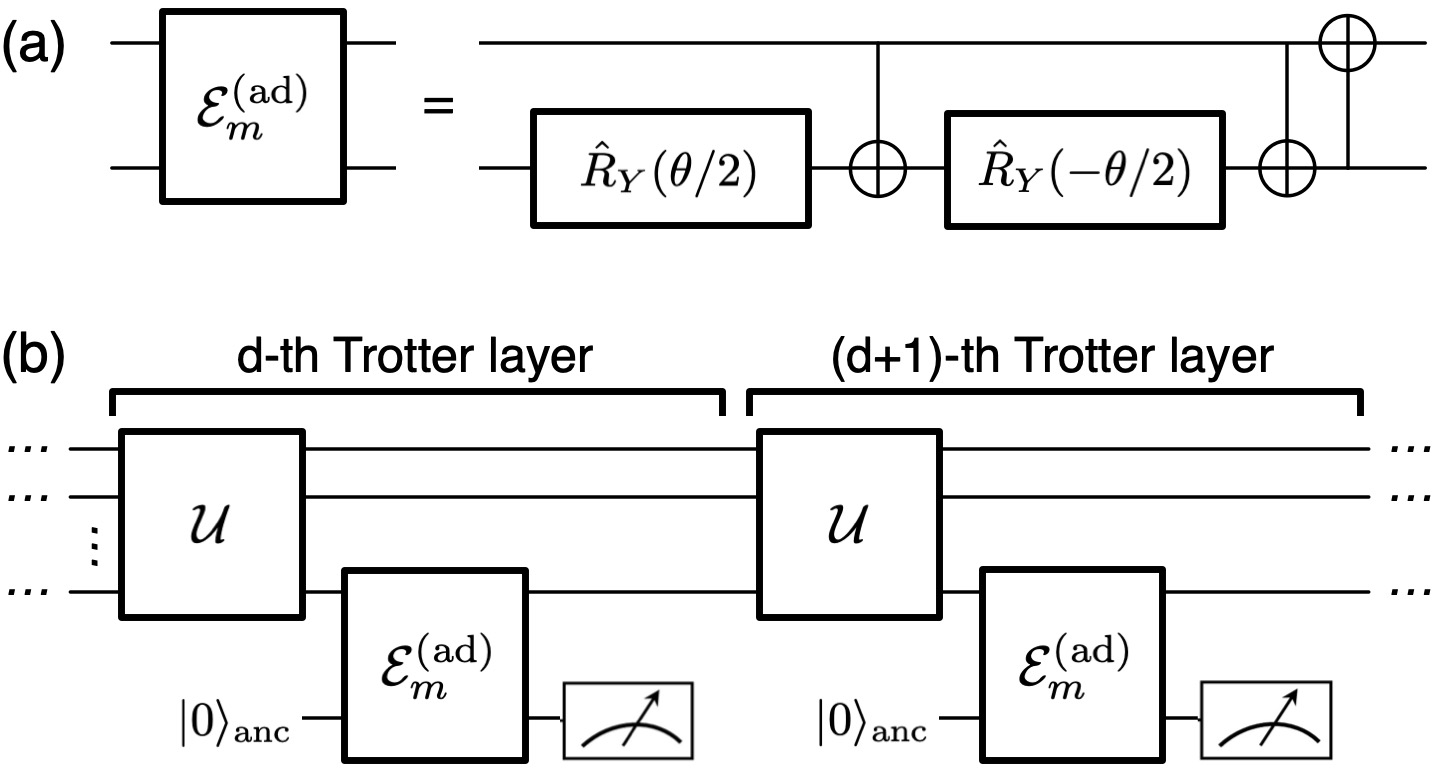}
 \caption{(a) Interaction $\mathcal{E}_{m}^{\rm (ad)}$ between open-system qubit $m$ and ancilla qubit, devised to introduce amplitude damping of the qubit $m$. (b) In each Trotter layer, the ancilla qubit is initialized in the state $\ket{0}$. Followed by the interaction $\mathcal{E}_{m}^{\rm (ad)}$ with qubit $m$, the ancilla qubit is reset via a measurement operation, so that it is reused in the next Trotter layer. The amplitude damping probability of qubit $m$ over a Trotter layer is $w_{m} = \sin^2(\theta/2)$.}
 \label{fig:amp_damp_circuit}
\end{figure}

The local amplitude damping of qubit $m$ is described by the Kraus operator in the form
\begin{equation}
    {\cal E}_{m}^{\rm (ad)}(\hat{\rho})= w_{m} \hat{\sigma}_{m}^{(-)}\hat{\rho}\hat{\sigma}_{m}^{(+)}+\hat{\sigma}_{m}^{(0)} \hat{\rho} \hat{\sigma}_{m}^{(0)},\label{eq:amplitude_damping}
\end{equation}
with $\hat{\sigma}_{m}^{(-)}=\ket{0}\bra{1}_{m}=\frac{1}{2}(\hat{X}_{m}+i\hat{Y}_{m})$, $\hat{\sigma}_{m}^{(+)}=\ket{1}\bra{0}_{m}=\frac{1}{2}(\hat{X}_{m}-i\hat{Y}_{m})$ and $\hat{\sigma}_{0}^{(m)}=\frac{1}{2}(1+\sqrt{1-w_{m}})\hat{I}_{m}+\frac{1}{2}(1-\sqrt{1-w_{m}})\hat{Z}_{m}$, where $w_{m}$ represents an incoherent transition probability from $\ket{1}$ to $\ket{0}$ of qubit $m$. The amplitude damping noise model in Eq.~\eqref{eq:amplitude_damping} cannot be described by the stochastic Pauli noise channels~\cite{blume2022taxonomy} in Eq.~\eqref{stochastic_channel}, thus requiring an approach different from Sec.~III. To that end, one could introduce an ancilla qubit coupled to the qubit $m$ where the interaction between them is described by a circuit shown in Fig.~\ref{fig:amp_damp_circuit}(a), consisting of three CNOT gates and single-qubit rotations $\hat{R}_{Y}(\pm\theta/2)=e^{\mp i\theta\hat{Y}_{a}/4}$ with a Pauli operator $\hat{Y}_{a}$ acting on the ancilla qubit \cite{nielsen2002quantum}. \jl{} In this approach, the initial state of the ancilla qubit is reset to $\ket{0}$ in each Trotter layer, as schematically shown in Fig.~\ref{fig:amp_damp_circuit}(b), and the amplitude damping probability $w_m=\sin^{2}(\theta/2)$ can be controlled by the single-qubit rotations $\hat{R}_{Y}(\pm\theta/2)$. The Lindblad dissipator corresponding to the repeated application of this quantum channel to each Trotter layer is described by
\begin{align}
    \mathcal{D}_{\rm ad}[\hat{\rho}] &=
    \sum_{m=1}^{n}\Gamma_m\left( \hat{\sigma}_{m}^{(-)}\hat{\rho} \hat{\sigma}_{m}^{(+)}-\frac{1}{2}\{\hat{\sigma}_{m}^{(+)}\hat{\sigma}_{m}^{(-)},\hat{\rho}\}\right), \label{eq:eq_AD_controlled}  \\
    \Gamma_m &= \text{ } w_{m}/\Delta t,\label{eq:gamma_AD_controlled}
\end{align}
where $\{\hat{A},\hat{B}\}$ denotes the anticommutator of two operators $\hat{A}$ and $\hat{B}$.

However, this approach has several drawbacks. The ancilla qubits are not desirable on current quantum hardware due to the limited number of qubits (up to $433$ qubits on the most recent IBMQ device), and the restricted qubit connectivity, e.g. superconducting devices have at most nearest-neighbour qubit connectivity distributed on a square 2D grid such as current Google quantum devices. Furthermore, additional two-qubit gates are needed to couple system qubits to ancilla qubits, as shown in Fig.~\ref{fig:amp_damp_circuit}, increasing the mitigation cost of our scheme. The use of mid-circuit measurement operations may also introduce non-negligible errors, both because of cross-talk and because the execution time of a measurement operation can be an order of magnitude longer than that of a two-qubit gate as in current IBMQ and Google superconducting quantum devices. During the long execution time of the measurement operation, unknown errors may occur, resulting in increased stochastic Pauli noise probabilities when Randomized Compiling is employed. Therefore, the measurement-based implementation of the amplitude damping may increase the total mitigation cost. To test this prediction, we performed simulations using the emulated quantum device \emph{ibmq oslo} and found that for an open system encoded in two qubits ($n=2$) the amplitude damping and stochastic Pauli noise channels can be implemented at the same time, but the total error probability $\sum_{k>0}\epsilon_k$ of the stochastic Pauli noise channels is substantially increased. This implies that the amplitude damping noise cannot be efficiently implemented by using the scheme in Fig.~\ref{fig:amp_damp_circuit}, based on ancilla qubits and measurement operations. In section \ref{sec: comparison_works} we comment on other recently proposed schemes to simulate amplitude damping which require ancilla qubits and face the same issues as the method previously discussed.

\subsection{Amplitude damping noise without ancilla qubits}

Here we propose a different approach where the amplitude damping is implemented solely by reset operations, which does not require ancilla qubits and mid-circuit measurement operations. 

We consider the stochastic application of the following reset channel to each Trotter layer
\begin{align}
    {\cal E}_{m}^{\rm (reset)}(\hat{\rho}) &= (1-w_{m})\hat{I}\hat{\rho}\hat{I}+w_{m}\mathcal{R}_{m}(\hat{\rho}),\label{eq:reset_AD}\\
    \mathcal{R}_{m}(\hat{\rho})&=\ket{0}\bra{0}_{m}\hat{\rho}\ket{0}\bra{0}_{m}+\hat{\sigma}_{m}^{(-)}\hat{\rho}\hat{\sigma}_{m}^{(+)},
\end{align}
where $\mathcal{R}_{m}(\hat{\rho})$ is a reset operation applied to qubit $m$ with probability $w_{m}$. Recently, fast high-fidelity reset operations (with reset failure probability $\sim 10^{-3}$ and time durations comparable to two-qubit gates with very low crosstalk error \cite{mcewen2021removing}) have been demonstrated for both superconducting \cite{mi2023stable} and ion-trap quantum devices \cite{chertkov2022holographic}. The stochastic application of the reset channel enables one to implement a Lindblad dissipator in the form
\begin{align}
    \mathcal{D}_{r}[\hat{\rho}] &= 
    \sum_{m=1}^{n}\Gamma_m\left(\hat{\sigma}_{m}^{(-)}\hat{\rho} \hat{\sigma}_{m}^{(+)}- \frac{1}{2}\{\hat{\sigma}_{m}^{(+)}\hat{\sigma}_{m}^{(-)},\hat{\rho}\}\right) \label{eq:eq_AD_Z_controlled} \\
    &\quad + \sum_{m=1}^{n}\frac{\Gamma_m }{4}( \hat{Z}_{m}\hat{\rho} \hat{Z}_{m} - \hat{\rho}),\nonumber\\
    \Gamma_{m} &= w_{m}/\Delta t,
\end{align}
which is a sum of the amplitude damping noise with rates $\Gamma_m$ and single-qubit dephasing with rates $\Gamma_m/4$. We note that the single-qubit dephasing rates can be controlled by our partial noise mitigation scheme in Sec.~\ref{sec:decoherence_control}. Therefore one can selectively implement the amplitude damping only, in principle, by using the reset operation.

\begin{figure}
\centering
\includegraphics[width=0.4\textwidth]{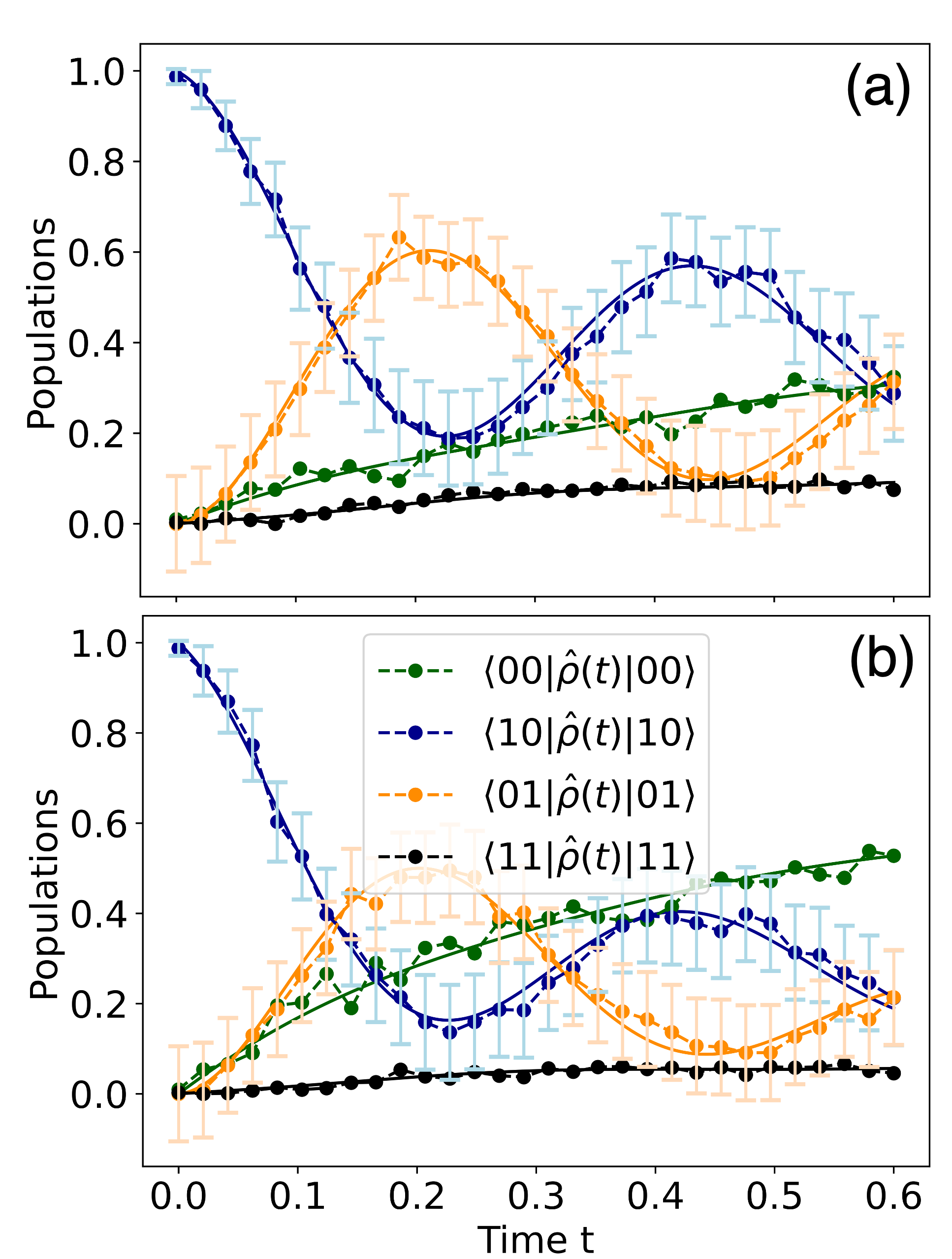}
\caption{Population dynamics of an open quantum system encoded in two qubits (n = 2), computed by the emulated
noisy IBMQ device (\emph{ibmq lagos}) with controlled stochastic Pauli noise and amplitude damping (dots), and solved classically by using the corresponding Lindblad equation (solid lines). Here we considered a uniform mitigation factor $r=0.2$ for the stochastic Pauli noise and an amplitude damping rate of (a) $J$ and (b) $3J$, with $J$ denoting the coupling strength between system qubits (see Eq.~\eqref{eq:H_PPC}). The single-qubit dephasing term arising from the reset channel in Eq.~\eqref{eq:eq_AD_Z_controlled} was fully mitigated. The Hamiltonian parameters considered in the simulation are as in Fig.~\ref{fig:evolve_normal}. We executed $90C_{\rm tot}^2$ quantum circuits for each data point.}
\label{fig:ad_controlled}
\end{figure}

To demonstrate that the amplitude damping noise can be efficiently implemented by the reset operation with controlled amplitude damping rates, we performed simulations on an emulated IBMQ device. After performing the noise characterization in Sec.~\ref{sec: noisy_evol}, we computed the time evolution of an open system encoded in two qubits with \emph{controlled} stochastic Pauli and amplitude damping noise. As shown in Fig.~\ref{fig:ad_controlled}, quantum simulation results are well matched to the classical solutions of the target Lindblad equation (see Eq.~\eqref{eq:D_stochastic_controlled} and \eqref{eq:eq_AD_controlled}). We assumed that the reset operation has an application time length $\sim 250$ $\rm ns$, similar to two-qubit gates, and reset failure probability of $\sim 10^{-3}$, which are consistent with the recent implementation of the reset operations on superconducting and ion-trap quantum hardware \cite{mi2023stable,mcewen2021removing,chertkov2022holographic}. We found that the error of the reset operation, considered in our simulations as $\mathcal{E}_{\rm er} = p_{\rm er}\hat{I}\hat{\rho}\hat{I}+(1-p_{\rm er})\mathcal{R}(\hat{\rho})$, being $p_{\rm er}\sim 10^{-3}$ the error failure probability, is comparable to that of single-qubit gates, which is one or two orders of magnitude lower than two-qubit gates. This error can also be accounted for in our simulations and mitigated by increasing the reset application probability $\omega_{m}$ in Eq. \eqref{eq:reset_AD}. This implies that a small number of reset operations applied to a quantum circuit does not require additional error mitigation and therefore can be employed to efficiently implement the amplitude damping model on current quantum devices. We note that the reset operations can be employed instead of the measurement operations in the ancilla-based scheme in Sec.~\ref{sec:amplitude_damping_with_ancilla} to avoid the errors caused by the measurements, but the mitigation cost increased by the ancilla qubits cannot be avoided.

\subsection{Generalized reset channels}

Here we show that other types of non-stochastic local noise can be implemented by using a combination of reset operations and single-qubit gates. We consider the following generalized reset channel
\begin{align}
    \mathcal{E}(\hat{\rho}) &= (1-p)\hat{I}\hat{\rho}\hat{I}+p\mathcal{V}(\hat{\rho}),\label{generalized_channel}\\
    \mathcal{V}(\hat{\rho}) &= \ket{\Psi}\bra{\Phi} \hat{\rho} \ket{\Phi}\bra{\Psi} + \ket{\Psi}\langle\Phi^{\perp}|\hat{\rho} |\Phi^{\perp}\rangle\bra{\Psi},
\end{align}
where $\ket{\Psi}=\hat{V}\ket{0}_{m}$, $\ket{\Phi}=\hat{U}\ket{0}_m$ and $|\Phi^{\perp}\rangle=\hat{U}\ket{1}_m$ with $\hat{U}^{\dag}$ and $\hat{V}$ denoting single-qubit gates applied before and after the reset operation acting on qubit $m$. The stochastic application of the generalized reset operation to each Trotter layer with probability $p$ leads to a Lindblad dissipator
\begin{equation}
    \mathcal{D}[\hat{\rho}] = \mathcal{D}_{\ket{\Psi}\bra{\Phi}}[\hat{\rho}]+\mathcal{D}_{\ket{\Psi}\bra{\Phi^{\perp}}}[\hat{\rho}],
\end{equation}
where
\begin{align}
\mathcal{D}_{\ket{\alpha}\bra{\beta}}[\hat{\rho}]&=\gamma\left(\ket{\alpha}\bra{\beta} \hat{\rho} \ket{\beta}\bra{\alpha}
    -\frac{1}{2}\{ \ket{\beta}\bra{\beta},\hat{\rho}\}\right), \\
    \gamma &= p/\Delta t.
\end{align}
As a special case, when $\hat{U} = \hat{V}$, the generalized reset channel leads to the amplitude damping and local dephasing in the $\{\ket{\Phi},|\Phi^{\perp}\rangle\}$ basis, described by
\begin{align}
    \mathcal{D}[\hat{\rho}] &= 
    \gamma \ket{\Phi}\langle \Phi^{\perp}|\hat{\rho} |\Phi^{\perp}\rangle\bra{\Phi}- \frac{\gamma}{2}\left\{|\Phi^{\perp}\rangle\langle \Phi^{\perp}|,\hat{\rho}\right\} \\
    &\quad + \frac{\gamma }{4}\left( \hat{U}\hat{Z}_{m}\hat{U}^{\dagger}\hat{\rho} \hat{U}\hat{Z}_{m}\hat{U}^{\dagger} - \hat{\rho} \right)\nonumber, 
\end{align}
where $\hat{U}\hat{Z}_{m}\hat{U}^{\dagger}=\ket{\Phi}\bra{\Phi}-|\Phi^{\perp}\rangle\langle \Phi^{\perp}|$. Therefore, by taking $\hat{U}\hat{Z}_{m}\hat{U}^{\dagger}$ to be a Pauli operator, such as $\hat{X}_m$, one can implement amplitude damping in the eigenbasis of the Pauli operator of choice, while the additional dephasing rate $\gamma/4$ is controlled by our partial noise mitigation scheme. In addition, one can implement the relaxation noise where the populations of $\ket{0}$ and $\ket{1}$ states are incoherently transferred to each other with different rates by stochastically applying the standard ($\hat{U}=\hat{I}$) and generalized ($\hat{U}=\hat{X}$) reset operations to each Trotter layer with different probabilities. Such a relaxation model, also known as generalized amplitude damping, has been widely considered in classical simulations of open system dynamics at finite temperatures~\cite{breuer2002theory}. More generally one can consider arbitrary single-qubit gates $\hat{U}$ and $\hat{V}$, which lead to various Lindblad dissipators that cannot be described by the stochastic Pauli noise models in Eq.~\eqref{eq:D_stochastic_controlled}.

\section{Comparison to previous works}\label{sec: comparison_works}

Several theoretical studies have proposed to introduce noise on quantum systems to improve the efficiency of quantum simulations, such as engineered dissipation for the preparation of ground states~\cite{cormick2013dissipative,raghunandan2020initialization,polla2021quantum} 
%and the use of the Zeno effect~\cite{geier2022non} \mbp{I removed this reference as I cannot see its particular relevance here}
(see recent reviews for more details, such as Ref.~\cite{harrington2022engineered}). In the context of quantum simulations of open system dynamics, analog quantum simulators provided the first step~\cite{mostame2012quantum,kim2022analog,lemmer2018trapped,gorman2018engineering,daley2022practical}. However, the programmability of these devices is typically restricted \cite{daley2022practical}, as one needs to map the Hamiltonian of interest to the ones that analogue quantum simulators can consider. %\jl{Comment: I removed the following sentence if you prefer.} Even though noise can be harnessed to aid analogue quantum simulations, a certain regime of noise may require one to replace experimental apparatus or quantum hardware \mbp{for example when the required noise rate is too high?} \jl{Answer: In general, the change of noise parameters in the quantum analogue processor may require the change of hardware components}, which may be experimentally burdensome to carry out. } 
In this context, leveraging the noise in digital quantum simulations has an advantage since one can consider a broader range of Hamiltonian models thanks to the universal set of quantum gates. Furthermore, our noise-assisted technique enables one to control the degree and type of Lindblad noise without changing quantum hardware at the cost of a classical overhead, namely the repetitions of quantum simulations. Therefore, both the open-system Hamiltonian and Lindblad noise are \emph{programmable} in noise-assisted digital quantum simulation.

Concerning previous proposals to digitally simulate open system dynamics, which do not harness the noise of quantum devices, a significantly larger number of qubits and two-qubit gates is required to take into account environmental degrees of freedom when compared to closed system simulations~\cite{barreiro2011open,lemmer2018trapped,cleve2016efficient,wang2011quantum,wang2023simulating,kamakari2022digital,schlimgen2021quantum,georgescu2014quantum,miessen2023quantum,nielsen2002quantum}. We remark that our noise-assisted technique does not require ancilla qubits and therefore additional two-qubit gates for the interaction between system and ancilla qubits, while its overhead comes in the form of a higher sampling cost, making it suitable for current NISQ devices. As the previous approaches do not use the noise as a resource, the full error mitigation is required to obtain accurate simulated results on NISQ devices. As discussed in Sec.~\ref{scalingPEC}, our partial noise mitigation scheme can provide an exponential speedup against the full noise mitigation considered in the previous approaches, namely a reduced classical overhead. Furthermore, we do not make use of variational approaches~\cite{kamakari2022digital} where resource estimation prior to quantum simulations is a non-trivial issue. As shown in Sec.~\ref{scalingPEC}, in our case, it is straightforward to estimate the computational resources required to simulate a given open system model on a specific quantum device.

To the best of our knowledge, there are only few theoretical studies that proposed to harness the noise of quantum devices for digital quantum simulations~\cite{rost2020simulation,leppakangas2022quantum,tolunay2023hamiltonian}. Our scheme has several advantages over previous approaches. Firstly, we do not assume a simple noise model solely described by $T_{1}$ and $T_{2}$ relaxation times~\cite{rost2020simulation,tolunay2023hamiltonian,leppakangas2022quantum}, namely amplitude damping and dephasing noise of individual qubits. Current NISQ devices are not fully characterized by the $T_{1}$ and $T_{2}$ relaxation times, since there are additional intrinsic noise channels, such as coherent and stochastic Pauli noise, including depolarizing channel~\cite{blume2022taxonomy,urbanek2021mitigating}. Contrary to Ref.~\cite{rost2020simulation,tolunay2023hamiltonian,leppakangas2022quantum}, we transformed the intrinsic noise of NISQ devices to the stochastic Pauli noise using Randomized Compiling and then performed noise characterization to fully take into account actual noise present in real quantum devices. Secondly, in the previous approaches, decoherence rates were controlled solely by changing a Trotter time-step $\Delta t$~\cite{rost2020simulation,leppakangas2022quantum,tolunay2023hamiltonian}. In this case, all the local amplitude damping and dephasing rates are uniformly modified in such a way that the ratio between decoherence rates is maintained. This implies that in case that the $T_{1}$ and $T_{2}$ relaxation times of qubits are inhomogeneous, as is the case of real NISQ devices, one cannot even implement a simple noise model of uniform decoherence rates by using the previous approaches. This is in contrast to our scheme where decoherence rates, including both stochastic Pauli noise and amplitude damping rates, can be controlled individually using partial noise mitigation and reset operations.

\begin{figure}
\centering
\includegraphics[width=0.45\textwidth]{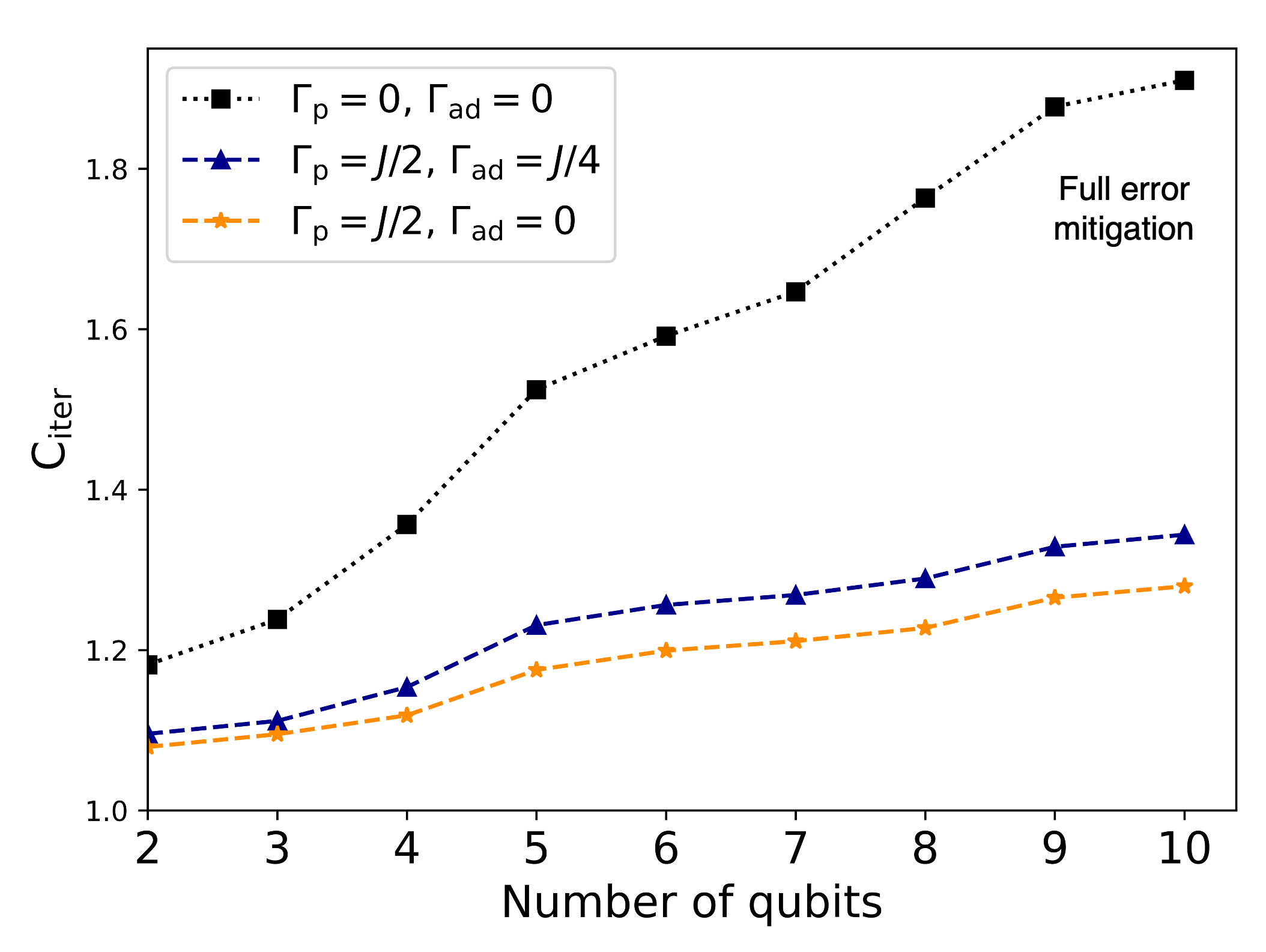}
\caption{Iteration mitigation cost $C_{\rm iter}$ of the TF Ising model as a function of the number $n$ of qubits, depending on uniform stochastic Pauli noise rate $\Gamma_p$ and amplitude damping rate $\Gamma_{\rm ad}$. The full noise mitigation results ($\Gamma_p = \Gamma_{\rm ad}=0$) are shown in black, while the partial noise mitigation results in the presence and absence of amplitude damping, namely $(\Gamma_p,\Gamma_{\rm ad})=(J/2,J/4)$ and $(\Gamma_p,\Gamma_{\rm ad})=(J/2,0)$, are shown in blue and orange, respectively. Hamiltonian parameters are taken to be $h=1$ and $J=0.5236$ with a Trotter time-step $\Delta t = 0.25$ as in Ref.~\cite{van2023probabilistic}. Noise characterization was performed on an emulated noisy IBMQ device (\emph{ibmq mumbai}).}
\label{cost_estimate}
\end{figure}

Recently, some of the techniques considered in our work, namely Randomized Compiling, Cycle Benchmarking and Probabilistic Error Cancellation, were employed to simulate a closed quantum system~\cite{van2023probabilistic} where all the stochastic Pauli noise was fully mitigated. The quantum system was encoded in ten qubits ($n=10$) and the Transverse Field (TF) Ising model was considered where the Hamiltonian is described by a one-dimensional array of qubits
\begin{equation}
    \hat{H} = -J\sum_{m=1}^{n-1}\hat{Z}_{m}\hat{Z}_{m+1} + h \sum_{m=1}^{n} \hat{X}_{m}.
\end{equation}
To demonstrate how the classical overhead of the full and partial noise mitigation schemes depends on the number of qubits, we consider single-qubit stochastic Pauli noise and amplitude damping channels, described by a Lindblad dissipator

\begin{align}
    \mathcal{D}[\hat{\rho}]&=\Gamma_{p}\sum_{m=1}^{n}\sum_{P\in\{X,Y,Z\}}(\hat{P}_m \hat{\rho}\hat{P}_m - \hat{\rho}) \\
    &\quad+ \Gamma_{\rm ad}\sum_{m=1}^{n}\left(\hat{\sigma}_{m}^{(-)}\hat{\rho}\hat{\sigma}_{m}^{(+)}-\frac{1}{2}\{\hat{\sigma}_{m}^{(+)}\hat{\sigma}_{m}^{(-)},\hat{\rho}\}\right),\nonumber
\end{align}

with uniform decoherence rates $\Gamma_{p}$ and $\Gamma_{\rm ad}$. In Fig.~\ref{cost_estimate}, we show the iteration mitigation cost $C_{\rm iter}=\prod_{m=1}^{n-1}C_{\rm mit}^{(m)}$ as a function of the number $n$ of qubits, from $n=2$ to 10, which determines how rapidly the total mitigation cost $\sim (C_{\rm iter})^{2D}$ increases with the number $D$ of Trotter layers. Note that $C_{\rm iter}$ of the partial noise mitigation scheme increases more slowly than that of the full mitigation scheme with $\Gamma_{p}=\Gamma_{\rm ad}=0$, implying that a larger number of qubits can be considered in open-system simulations than in closed-system simulations for a given total mitigation cost. For a given set of target decoherence rates of the stochastic Pauli noise channels, $C_{\rm iter}$ becomes larger when the amplitude damping channel is included, as shown in Fig.~\ref{cost_estimate}, as the reset operations increase the local dephasing rates (see Eq.~\eqref{eq:eq_AD_Z_controlled}). These results demonstrate that our noise-control technique may be suitable for quantum simulations of large open-system dynamics under Lindblad noise on current NISQ devices, which deserves a separate investigation and will be presented in a forthcoming manuscript.

\section{Conclusions} \label{conclusions}
In this work, we have established the concept of intrinsic noise-assisted digital quantum algorithms and demonstrated the 
principle in real and emulated IBM Quantum computers based on superconducting qubits.
We have shown that the application of intrinsic noise-assisted digital quantum algorithms in real world devices can be 
achieved using three key steps. First, the coherent system dynamics is decomposed via Trotterization into product formulae with
time-step $\Delta t$ of our choice to gain a first level of control over the effective noise realised in the quantum 
simulation. Based on this, the second crucial step is the characterization of the intrinsic noise in the NISQ device on which the algorithm is run, which depends on how the system Hamiltonian is implemented within a Trotter layer. This can be achieved via Randomized Compiling, Cycle Benchmarking and Error
Reconstruction techniques. Building on this intrinsic noise reconstruction, the third and final step uses Probabilistic Error
Cancellation to control the noise in the quantum circuit, enabling implementation of a target decoherence model, including both stochastic and generalized amplitude damping noise based on reset operations.

The principle of digital quantum simulation assisted by the device's intrinsic noise offers
a number of advantages in the field of NISQ computation. First, the noise-assisted digital 
quantum algorithm does not need to alter the quantum hardware to tune the intrinsic noise but achieves it via
results postprocessing. Secondly, it does not require additional {\it quantum} computational resources, i.e. additional 
qubits and CNOT gates, for simulating open quantum systems compared to more standard approaches~\cite{barreiro2011open,hu2022general,cleve2016efficient,wang2011quantum,kamakari2022digital,schlimgen2021quantum}.
The quantum resource reduction is achieved via an additional {\it classical} overhead, i.e. an increased number
of runs of the algorithm in the device, due to the use of a quantum error mitigation technique to control the noise acting on the qubits. As a result, our work provides guiding principles for the execution of
quantum digital simulation of open quantum systems on real world devices, where noise is not detrimental, but 
leveraged for a more efficient computation. 

In future work, we expect that the results can be extended in a variety of fruitful directions. First of all, it will
be interesting to benchmark (and adapt) the proposed noise-assisted technique on (to) different quantum hardware technologies beyond 
superconducting quantum devices which may exhibit particularly useful intrinsic noise. Secondly, in this spirit, 
the addition of modest quantum resources may allow for the extension to the efficient digital simulation of 
non-Markovian environments combining it with techniques exposed in Ref.~\cite{lemmer2018trapped,mascherpa2020optimized}. For instance, in the case of bosonic environments, it has been shown that a continuous bosonic environment interacting with an open system can be effectively described by a few bosonic modes under Lindblad noise, such as pseudo and surrogate modes, when an optimal Lindblad noise model is constructed based on environmental correlation functions~\cite{lemmer2018trapped,mascherpa2020optimized}. In case that such a concept of an effective small environment under Lindblad noise is considered in quantum simulations, one needs to find a way to control the noise on quantum devices and in this context our noise-control technique may be useful.
Thirdly, we expect these techniques to find fruitful combinations with Quantum TEDOPA~\cite{guimaraes2022efficient}, a technique to simulate non-perturbative dynamics of open quantum systems in a 
quantum computer, by making use of the results of a recent work on Markovian closure~\cite{nusseler2022fingerprint}.
Finally, we expect further efficiency enhancements of our technique to be possible with improvements in tailormade
characterization and error control techniques to give access to general non-perturbative dynamics of open quantum systems.

\section*{Acknowledgements} The authors acknowledge helpful discussions and support by Benjamin Desef. JDG acknowledges funding from the 
Portuguese Foundation for Science and Technology (FCT) through PhD grant UI/BD/151173/2021. JDG, JL, MBP acknowledge support by the BMBF 
project PhoQuant (grant no. 13N16110). MIV acknowledges support from the FCT through Strategic Funding UIDB/04650/2020. SFH and MBP acknowledge
support by the DFG via the QuantERA project ExtraQt. The authors acknowledge support by the state of Baden-Württemberg through bwHPC and the 
German Research Foundation (DFG) through Grant No. INST 40/575-1 FUGG (JUSTUS 2 cluster) and acknowledge the use of IBMQ devices via the 
Researcher Program.

\appendix

\section{Randomized Compiling}\label{A}

The noise we considered in the quantum simulations (see Sec.~\ref{sec: noisy_evol}) is given by stochastic Pauli noise channels defined as $\sum_{k}\epsilon_{k}\hat{P}_{k}\hat{\rho}\hat{P}_{k}$. The absence of coherent noise $\sum_{k'\neq k}\hat{P}_{k}\hat{\rho}\hat{P}_{k'}$ in this stochastic noise model allows us to efficiently cancel it in the quantum circuit by implementing Pauli operators via the PEC scheme (see Sec.~\ref{control_decoh}). Therefore, a stochastic Pauli channel is a desirable noise model to have in quantum circuits since it can be straightforwardly cancelled by PEC. In superconducting quantum devices however, several types of noise may occur on the course of the implementation of a set of gates, specifically coherent errors arising from qubit cross-talk or under/over-rotations of gates, and incoherent ones, such as amplitude damping or stochastic Pauli noise. In view of obtaining a stochastic noise model in our quantum simulations, we applied a quantum error mitigation technique, namely Randomized Compiling \cite{wallman2016noise,hashim2020randomized}, that transforms coherent noise into stochastic Pauli noise, both in the noise characterization routines and in the digital quantum simulations implemented on the real IBMQ devices. 

\begin{figure}[b]
\centering
\includegraphics[width=0.4\textwidth]{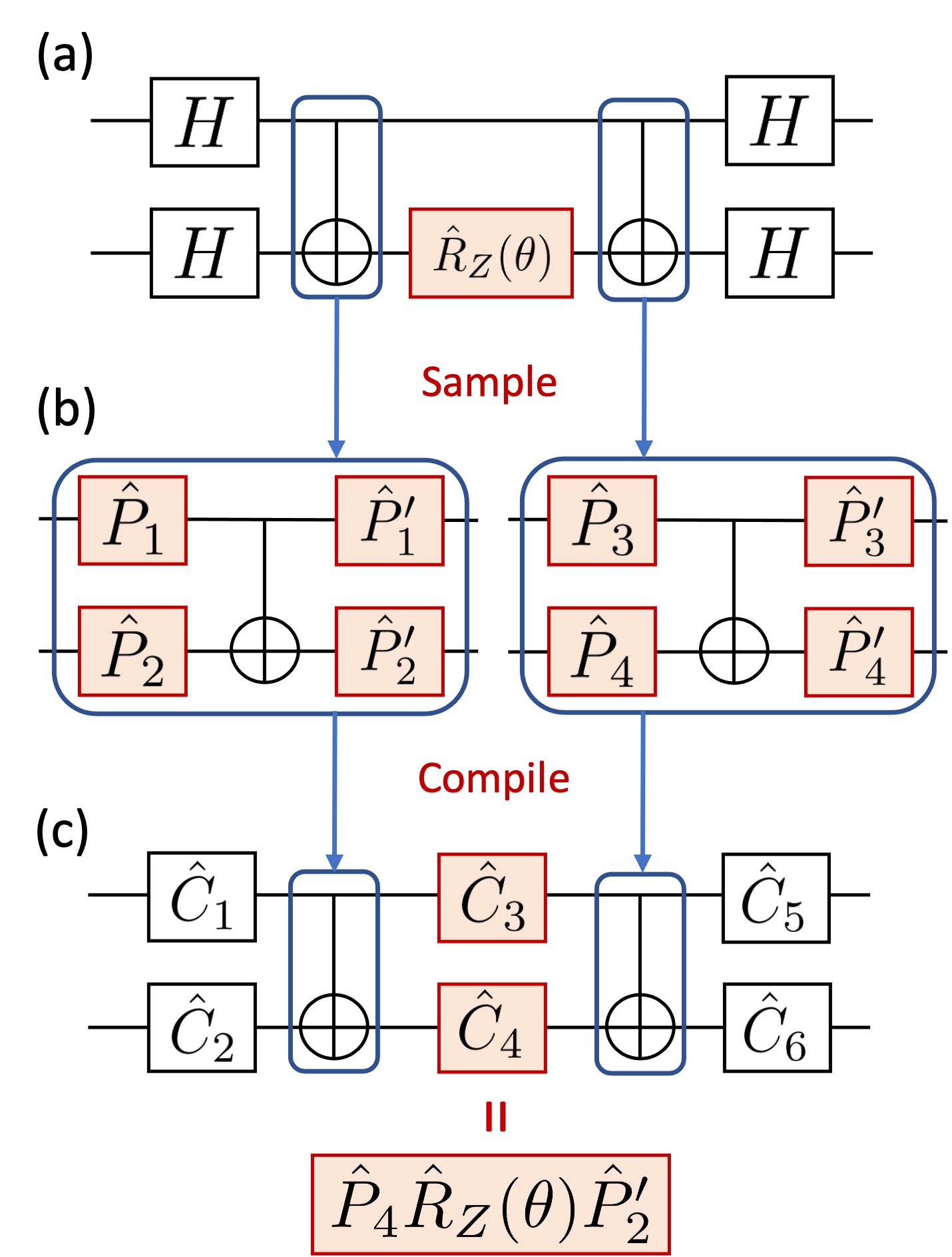}
\caption{Example of the application of Randomized Compiling to the operation $e^{-i\frac{\theta}{2} \hat{X}\otimes \hat{X}}$, a term that appears when evolving a quantum system via the Hamiltonian in Eq.~\eqref{eq:H_PPC}, decomposed into Hadamard ($H$), CNOT and $\hat{R}_{Z}(\theta)$ gates as shown in circuit (a). Randomized Compiling consists of, firstly applying uniformly sampled random two-qubit Pauli strings $\hat{P}_{k}$ to the CNOT gates in circuit (a) together with the Pauli gates $\hat{P}'_{k}$ as defined in Eq.~\eqref{RC_eq}. This procedure is displayed in circuit (b). The next and final step is to compile the circuit, such that the added Pauli strings in the circuit are absorbed into other nearby single-qubit quantum gates. This compilation process creates a new single-qubit gate $\hat{C}_{k}$ from the previous ones as shown in circuit (c).}
\label{fig:RC}
\end{figure}

 Randomized Compiling consists of creating several copies of the circuit and for each noisy gate $\hat{G}$ acting on $Q$ qubits in each copy, a randomly sampled $Q$-qubit Pauli string $\hat{P}_{k}$ is applied as follows, 
\begin{equation}
    \hat{G} = \hat{P'}_{k}\hat{G} \hat{P}_{k}, \quad \hat{P'}_{k}=\hat{G} \hat{P}_{k}\hat{G}^{\dag}. \label{RC_eq}
\end{equation}
We assumed $\hat{G}$ to be part of the Clifford group, such that $\hat{P'_{k}}$ is another Pauli string. Note that if $\hat{G}$ is not part of the Clifford group, $\hat{P'}$ would not be a Pauli string, and potentially additional two-qubit gates would be required to be applied in the circuit \cite{hashim2020randomized,wallman2016noise}, hence increasing error probabilities.

CNOT gates are the most noisy gates in a quantum circuit, hence we applied Randomized Compiling to them as exemplified in Fig.~\ref{fig:RC}. We executed several $R$ randomized compiled circuits and then classically averaged the outcomes, thus obtaining an average stochastic Pauli noise process acting on the qubits.

\begin{figure}
\centering
\includegraphics[width=.48\textwidth]{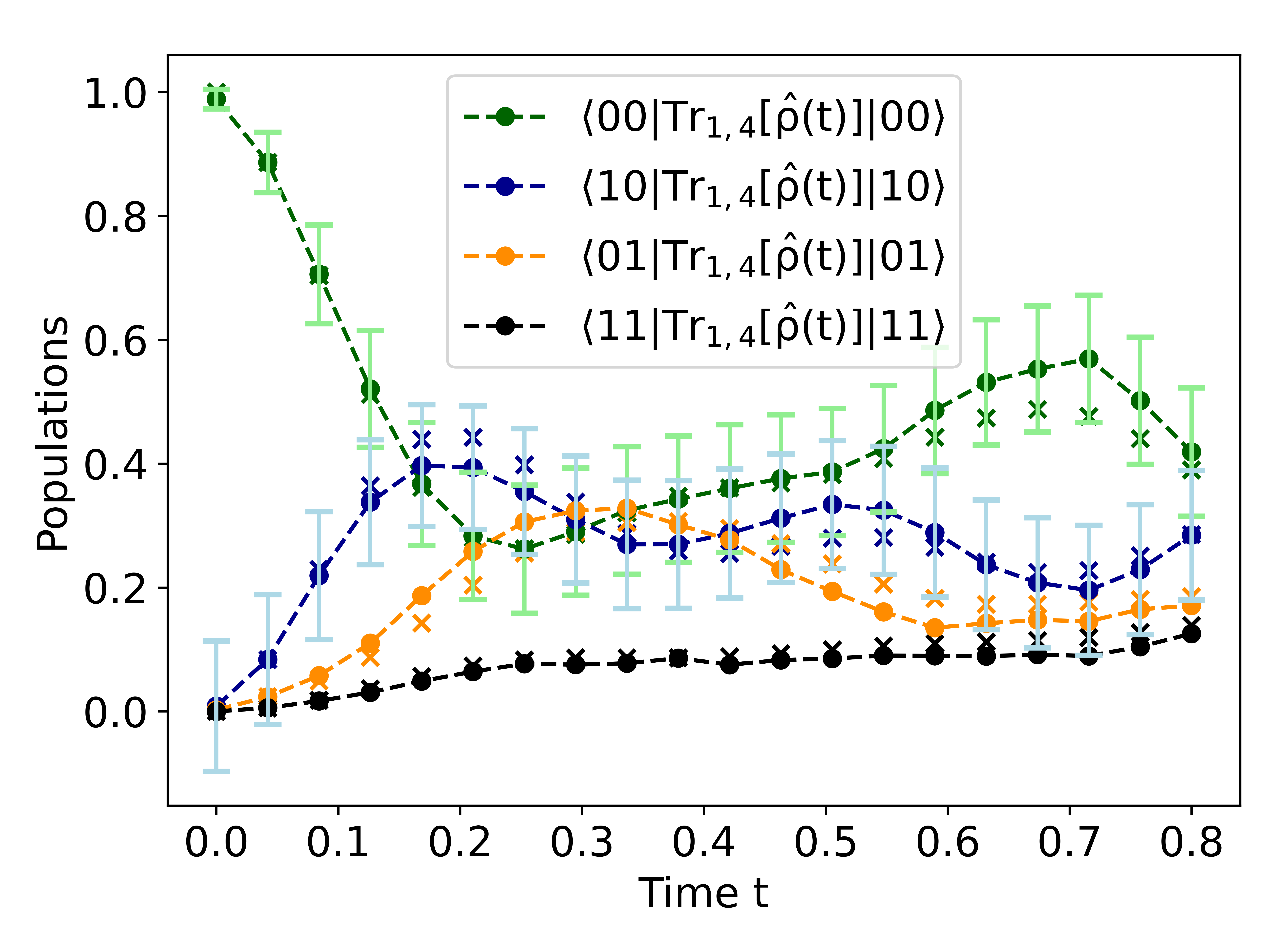}
\caption{Population dynamics of an open system encoded in four qubits ($n=4$) simulated by a quantum computer with non-uniform mitigation factors, shown in dots, which are well matched to classical solutions of the corresponding Lindblad equation, shown in crosses. Here the time evolution of the reduced density matrix ${\rm Tr}_{1,4}[\hat{\rho}(t)]$ of the second and third qubits is displayed. The one- and two-qubit dephasing noise channels, namely $\hat{P}_{k}\in\{\hat{Z}_{m},\hat{Z}_{m}\hat{Z}_{m+1}\}$, were mitigated by $r_{k}=0.5$, while all the other $K=2$ stochastic Pauli noise channels $\hat{P}_{k'}$ were weakly mitigated by $r_{k'}=0.1$. Hamiltonian parameters and initial state are as in Fig.~\ref{fig:evolve_normal}, and $90 C_{\rm tot}^{2}$ circuits were considered in PEC.}
\label{fig:evolve_PEC_4qubs}
\end{figure}

\section{4-qubit noise-controlled evolution simulations with interaction-specific mitigation}\label{D}
We performed a noise characterization and noise-controlled quantum simulation of an 1D array of $n=4$ qubits by characterizing and mitigating two-qubit stochastic Pauli channels (i.e. $K=2$) on the emulated \emph{ibmq lagos} device. The structure of the circuit is shown in Fig.~\ref{fig:trotter_circuit}(b), where we used a first-order Trotter-Suzuki product formula. We implemented non-uniform mitigation factors in the quantum simulations and compared the results with a classically solved Trotterized Lindblad equation where the decoherence rates are given by equation \eqref{eq:gamma_stochastic_controlled}. The results are shown in Fig. \ref{fig:evolve_PEC_4qubs}. 
 We measured the population terms of the reduced two-qubit density matrix (second and third qubit in the 1D array) on the emulated IBMQ device. 
 
 The quantum simulation results show a good qualitative agreement with the classical simulation outcomes, suggesting that the proposed noise-assisted simulation technique with non-uniform mitigation factors can be applied for $n>K=2$ digital quantum simulations on IBMQ computers.

\bibliography{apssamp}

%\begin{thebibliography}{99}

%\bibitem{yuri1980computable} Y. Manin, Computable and Uncomputable, Sov. Radio. {\bf 128}, ### (1980).

%\bibitem{feynman1982simulating} R. P. Feynman, Simulating Physics with Computers, Int. J. Theor. Phys. {\bf 21}, 467 (1982).

%\bibitem{breuer2002theory} H.-P. Breuer and F. Petruccione, The Theory of Open Quantum Systems (Oxford University Press, 2002).

%\bibitem{rivas2012open} A. Rivas and S. F. Huelga, Open Quantum Systems: An Introduction (Springer, 2012).

%\bibitem{tanimura1989time} Y. Tanimura and R. Kubo, Time Evolution of a Quantum System in Contact with a Nearly Gaussian-Markoffian Noise Bath, J. Phys. Soc. Jpn. {\bf 58}, 101 (1989).

%\bibitem{prior2010efficient} J. Prior, A. W. Chin, S. F. Huelga, and M. B. Plenio, Efficient Simulation of Strong System-Environment Interactions, Phys. Rev. Lett. {\bf 105}, 050404 (2010).

%\end{thebibliography}

\end{document}